\documentclass[aps,prd,reprint,superscriptaddress,floatfix,nofootinbib]{revtex4-2}
\usepackage[T1]{fontenc}
\usepackage{lmodern}
\usepackage[utf8]{inputenc}
\usepackage{mathtools}
\usepackage{graphicx}
\usepackage{xcolor}
\usepackage{amsmath,latexsym}
\usepackage{eurosym}
\usepackage{amsfonts}
\usepackage[hidelinks]{hyperref}
\usepackage{orcidlink}

\hypersetup{
  pdftitle={Charged Letelier–Alencar Black Strings with Quintessence and the Cosmological Constant},
  pdfauthor={B. Eslam Panah, F. M. da Silva, and L. G. Barbosa}
}

\newcommand{\jcap}{JCAP}
\newcommand{\apss}{Astrophys. Space Sci.}

\begin{document}

\title{Charged Letelier--Alencar Black Strings with Quintessence and the Cosmological Constant}
\author{B. Eslam Panah\,\orcidlink{0000-0002-1447-3760}}
\email{eslampanah@umz.ac.ir}
\affiliation{Department of Theoretical Physics, Faculty of Basic Sciences, University of
Mazandaran, P. O. Box 47416-95447, Babolsar, Iran}
\affiliation{Center for Theoretical Physics, Khazar University, 41 Mehseti Str., Baku,
AZ1096, Azerbaijan}
\author{F. M. da Silva \orcidlink{0000-0003-2568-2901}}
\email{franciele.m.s@ufsc.br}
\affiliation{Departamento de F\'isica, CFM - Universidade Federal de Santa Catarina, \\
Caixa Postal 5064, CEP 880.35-972, Florian\'opolis, SC, Brazil.}
\affiliation{Theoretical Astrophysics, Institute for Astronomy and Astrophysics,
University of T\"{u}bingen, 72076 T\"{u}bingen, Germany}
\author{L. G. Barbosa \orcidlink{0009-0007-3468-3718}}
\email{leonardo.barbosa@posgrad.ufsc.br}
\affiliation{Departamento de F\'isica, CFM - Universidade Federal de Santa Catarina, \\
Caixa Postal 5064, CEP 880.35-972, Florian\'opolis, SC, Brazil.}

\begin{abstract}
We obtain a new exact solution for a charged black string coupled
to a Letelier--Alencar cloud of strings and a Kiselev fluid in
Einstein--Maxwell--$\Lambda$ theory. We study the curvature scalars in order to investigate the asymptotic behavior of these solutions. We calculate the relevant conserved and thermodynamic quantities, confirming their compliance with the first law of thermodynamics. System stability is rigorously analyzed through heat capacity and Gibbs free energy, identifying regimes of local and global equilibrium. Additionally, by interpreting the cosmological constant as thermodynamic pressure within an extended phase space, we establish the validity of the extended first law and derive the corresponding Smarr relation for these solutions. We also investigate the null geodesic structure, deriving the
equation for the photon cylinder radius, which generalizes the known result for charged black strings by incorporating the effects of the
Letelier--Alencar cloud of strings and quintessence. The numerical analysis of this equation determines the existence and location of
photon cylinders as functions of the model parameters.
\end{abstract}

\maketitle

\section{Introduction}

\label{Introduction}

Black strings, first introduced by Lemos \cite{Lemos:1994xp}, are exact cylindrical solutions of the Einstein field equations with a negative cosmological constant. Unlike standard black holes with spherical horizons, black strings possess horizons that extend along an infinite spatial direction, giving them a cylindrical topology. The static solution was later extended to include electric charge and rotation by Lemos and Zanchin \cite{Lemos:1995cm}. The stability \cite{Gregory:1993vy,Gregory:1994bj,Yoo:2011vu}, thermodynamical properties \cite{Fatima2012Ap&SS,Ahmed:2025sav} and Hawking radiation \cite{Ahmed2011JCAP,Ahmed2011JCAP2} of the black strings have been studied in many works, as well as its connection to other geometries \cite{Lima2023Symm,Jusufi:2022rbt}. Additionally, these configurations have been studied in various contexts, including modified gravity \cite{Tannukij2017EPJC,Cisterna:2018jsx,Darlla:2023qgf,Santos:2026bjq} and matter configurations \cite{Ali:2019mxs, Deglmann:2025mcl,Barbosa2026PhLB,Barbosa2026arXiv260606435B,Barbosa:2025scy,Cunha:2022kep}.

The influence of anisotropic matter distributions on gravitational solutions has also received considerable attention. A prominent example is the cloud of strings introduced by Letelier \cite{Letelier:1979ej}, which describes a continuous distribution of one-dimensional strings through an anisotropic energy--momentum tensor. Another widely studied source is the anisotropic fluid proposed by Kiselev \cite{Kiselev:2002dx}, which provides an effective description of quintessence surrounding compact objects by means of a barotropic equation of state parameter. These matter sources have been extensively employed to investigate how string-like matter and dark-energy environments modify the spacetime geometry, horizon structure, geodesic motion, and thermodynamic properties of black hole solutions.

In 2025, Alencar et al. introduced a generalized configuration of a cloud of strings \cite{Alencar}. In this framework, the original Letelier cloud of strings model \cite{Letelier:1979ej} is extended by incorporating an additional magnetic-like component, denoted by $\Sigma_{23}$. The presence of this term leads to a fully anisotropic energy--momentum tensor of the form $\mathrm{diag}(-\rho,-\rho,p,p)$, which is characterized by two
parameters and governed by a single equation of state. This extension
enriches the matter sector while preserving the fundamental geometrical features of the original model, namely spherical symmetry and anisotropy. Consequently, the generalized model provides a broader and more flexible framework for investigating the influence of extended string-like matter distributions on the spacetime geometry, horizon structure, and thermodynamic properties of black holes \cite{Alencar}. In this regard, relativistic tidal forces and geodesic motion around a black hole surrounded by a generalized Letelier--Alencar cloud of strings have been investigated in Ref. \cite{Silva2026}. How the cloud of string parameters, $g_s$ and $l_s$, modify the curvature singularity, photon sphere, innermost stable circular orbits, and the compression/stretching profiles of tidal forces for both radially infalling and circular observers has also been analyzed. Black hole solutions in Einstein gravity coupled to a $U(1)$ gauge field, a Letelier--Alencar cloud of strings, and a cosmological constant are obtained, and their horizon structure and thermodynamic quantities are examined \cite{EslamPanah2026}. Local and global thermodynamic stability are analyzed through the heat capacity and Gibbs potential, while the model parameters are constrained using Bayesian MCMC fits to quasi-periodic oscillation data from black holes across different mass scales. In Ref. \cite{Muniz2026PDU}, the authors considered the Letelier--Alencar cloud of strings to construct a family of regular black holes solutions and studied their properties, calculating the shadow radius and imposing constraints consistent with the Event Horizon Telescope bounds for Sgr A* and M87*.

Conversely, electrically charged black strings in the presence of the
cosmological constant offer a compelling paradigm for the study of black string thermodynamics. The presence of a negative cosmological constant not only reshapes the asymptotic structure of the spacetime but also induces nontrivial phase transitions and stable thermodynamic states. By adopting the extended phase-space formalism, wherein the cosmological constant is identified as a thermodynamic pressure \cite{PLambda2,PLambda3,PLambda4},
one can examine the first law of thermodynamics and the associated Smarr relation \cite{Smarr1973PhRvL,Smarr1973PhRvL1} within a more rigorous and comprehensive framework. Furthermore, the inclusion of electric charge extends the system's thermodynamic degrees
of freedom, thereby significantly deepening the analysis of its phase
structure and stability criteria.

Motivated by the preceding discussion, charged Letelier--Alencar black strings constitute a physically well-motivated framework for investigating the interplay among electromagnetic effects, the cosmological constant, and generalized cloud of strings matter. Within this setting, one can systematically examine how the modified matter sector (particularly the magnetic-like contribution and the anisotropic pressure induced by it) affects the underlying gravitational configuration and the principal physical properties of the solution. In particular, this model provides a suitable basis for analyzing the horizon structure, conserved quantities, thermodynamic response, local and global stability conditions, and the extended phase space behavior of the system. Accordingly, the present work aims to elucidate the role of nontrivial string-like matter distributions in shaping the geometry and thermodynamic dynamics of black string spacetimes.

In this work, we solve the Einstein--Maxwell--$\Lambda$ field equations for a charged black string coupled to a Letelier--Alencar cloud of strings and quintessence. We analyze the horizon structure, curvature scalars, and thermodynamic properties of the solution. We also investigate the null geodesic structure, deriving the equation for the photon cylinder radius. The solution generalizes known charged black string spacetimes by simultaneously incorporating the generalized cloud of strings and quintessence.

The paper is organized as follows. In Sec. \ref{Action_and_Field_Equations} we present the field equations and derive the solution. In Sec. \ref{Solution_and_Curvature_Scalars} we analyze the curvature scalars. In Sec. \ref{Thermodynamic} we study the thermodynamic quantities, stability, and extended phase space. In Sec. \ref{Photon Cylinder} we investigate the photon cylinder. Finally, in Sec.~\ref{Conclusions} we present our conclusions.

\section{Field Equations}

\label{Action_and_Field_Equations}

The Einstein--Maxwell field equations with a cloud of strings (specifically,
the Letelier--Alencar model), a Kiselev fluid, and the cosmological constant are given by
\begin{eqnarray}
G_{\mu }^{~\nu }+\Lambda \delta_{\mu}^{\nu}{+}2\left( \frac{\delta_{\mu}^{\nu}%
\mathcal{F}}{4}-F_{\mu }^{~\alpha }F_{~\alpha }^{\nu }\right) &=&8\pi \left(
T_{\mu }^{~\nu ^{CS}}+T_{\mu }^{~\nu ^{Q}}\right) ,  \label{FE} \\
&&  \notag \\
\partial_{\mu}\left( \sqrt{-g}F^{\mu \nu }\right) &=&0,  \label{Ftr}
\end{eqnarray}%
where $G_{\mu }^{~\nu }=\mathcal{R}_{\mu }^{~\nu }-\frac{1}{2}\delta_{\mu}^{\nu}\mathcal{R}$ and $\mathcal{R}$ is the Ricci scalar. In Eq. (\ref{FE}), $\Lambda$ is the cosmological constant and $\delta_{\mu}^{\nu}$ is the Kronecker delta. Also, $\mathcal{F}=F_{\mu \nu }F^{\mu \nu }$,
where $F_{\mu \nu }=\partial _{\mu }A_{\nu }-\partial _{\nu }A_{\mu }$ is
the Faraday tensor with $A_{\mu }$ as the gauge potential. In Eq. (\ref{Ftr}%
), $g$ refers to the determinant of the metric tensor $g_{\mu \nu }$, i.e., $%
g=\det \left( g_{\mu \nu }\right) $. It is notable that we set $c=G=1$ throughout this paper.

In Eq. (\ref{FE}), $T_{\mu }^{~\nu ^{CS}}$ is the energy-momentum tensor of
the Letelier--Alencar cloud of strings in the following form \cite{Alencar} 
{\setlength{\arraycolsep}{.5pt}\small
\begin{equation}
T_{\mu }^{~\nu ^{CS}}=\left( 
\begin{array}{cccc}
\frac{-g_{s}^{2}\sqrt{l_{s}^{4}+r^{4}}}{8\pi r^{4}} & 0 & 0 & 0 \\ 
0 & \frac{-g_{s}^{2}\sqrt{l_{s}^{4}+r^{4}}}{8\pi r^{4}} & 0 & 0 \\ 
0 & 0 & \frac{g_{s}^{2}l_{s}^{4}}{8\pi r^{4}\sqrt{l_{s}^{4}+r^{4}}} & 0 \\ 
0 & 0 & 0 & \frac{g_{s}^{2}l_{s}^{4}}{8\pi r^{4}\sqrt{l_{s}^{4}+r^{4}}}%
\end{array}%
\right) ,  \label{TCS}
\end{equation}
}%
where $l_{s}$ and $g_{s}$ are the string length and the effective
coupling controlling the gravitational contribution of the cloud of strings,
respectively \cite{Alencar}. It is notable that the components of $T_{\mu
}^{~\nu ^{CS}}$ reduce to Letelier cloud of strings in the limit $%
l_{s}\rightarrow 0$ and $g_{s}^{2}\rightarrow \alpha_{\mathrm{CS}}$. Here $\alpha_{\mathrm{CS}}$ denotes the amplitude of the Letelier cloud.

In Eq. (\ref{FE}), $T_{\mu }^{~\nu ^{Q}}$ is the energy-momentum tensor of
the quintessence in the following form \cite{Kiselev:2002dx} 
{\setlength{\arraycolsep}{.5pt}\small
\begin{equation}
T_{\mu }^{~\nu ^{Q}}=\left( 
\begin{array}{cccc}
{-}\frac{3\omega \sigma }{8\pi r^{3\left( \omega +1\right) }} & 0 & 0 & 0 \\ 
0 & {-}\frac{3\omega \sigma }{8\pi r^{3\left( \omega +1\right) }} & 0 & 0 \\ 
0 & 0 & \frac{3\omega \sigma \left( 3\omega +1\right) }{16\pi r^{3\left(
\omega +1\right) }} & 0 \\ 
0 & 0 & 0 & \frac{3\omega \sigma \left( 3\omega +1\right) }{16\pi r^{3\left(
\omega +1\right) }}%
\end{array}%
\right) ,  \label{TQ}
\end{equation}
}%
where $\omega$ represents the state parameter of the quintessence field restricted to the range $-1 < \omega < -1/3$, and $\sigma$ denotes the corresponding normalization constant. It is worth noting that the contribution of the quintessence field vanishes in the limit $\sigma = 0$. The two anisotropic stress tensors in Eqs.~(\ref{TCS}) and (\ref{TQ}) are used as effective matter sources in the cylindrical ansatz.

We consider a four-dimensional static spacetime in the following form \cite%
{Lemos:1994xp} 
\begin{equation}
ds^{2}=-f\left( r\right) dt^{2}+\frac{dr^{2}}{f\left( r\right) }%
+r^{2}d\varphi ^{2}+r^{2}\alpha ^{2}dz^{2},  \label{metric}
\end{equation}%
where $f\left( r\right) $ is the metric function.  Also, $\alpha$ is a constant parameter with dimensions of inverse length ($[\alpha]=L^{-1}$),
associated with the characteristic AdS curvature radius. For $\Lambda<0$, we use the axial normalization $\alpha=\sqrt{-\Lambda/3}$ of the AdS black string. Also, the coordinates
ranging as $-\infty <t<+\infty $, $0\leq r<+\infty $, $0\leq \varphi \leq
2\pi $, and $-\infty <z<+\infty $.

Now, we consider a radial electric field, whose associated gauge potential
is $A_{\mu }=h\left( r\right) \delta _{\mu }^{t}$. Using Eqs. (\ref{Ftr}), (%
\ref{metric}) and the ansatz for $A_{\mu }$, we obtain the following differential equations 
\begin{equation}
2h^{\prime }\left( r\right) +rh^{\prime \prime }\left( r\right) =0,
\end{equation}%
where prime and double prime, respectively, denote the first and second
derivations with respect to the radial coordinate. We find $h(r)$ as 
\begin{equation}
h\left( r\right) =\frac{-2q}{r},  \label{h(r)}
\end{equation}%
where $q$ is an integration constant related to the electric charge.

By employing Eqs. (\ref{TCS}), (\ref{TQ}), (\ref{metric}), and Eq. (\ref%
{h(r)}) within Eq. (\ref{FE}), we can find the components of the equations
of motion (Eq. (\ref{FE})), which are 
\begin{widetext}
\begin{eqnarray}
eq_{tt} &=&eq_{rr}=\Lambda r^{2}+rf^{\prime }\left( r\right) +f\left(
r\right) +\frac{4q^{2}}{r^{2}}+\frac{g_{s}^{2}\sqrt{l_{s}^{4}+r^{4}}}{r^{2}}+%
\frac{{3\omega \sigma }}{r^{3\omega +1}}=0,  \label{eqtt} \\
&&  \notag \\
eq_{\varphi \varphi } &=&eq_{zz}=\left( f^{\prime \prime }\left( r\right)
+2\Lambda \right) r^{2}+2rf^{\prime }\left( r\right) -\frac{8q^{2}}{r^{2}}{-}%
\frac{9\omega \sigma \left( \omega +\frac{1}{3}\right) }{r^{3\omega +1}}-%
\frac{2g_{s}^{2}l_{s}^{4}}{r^{2}\sqrt{l_{s}^{4}+r^{4}}}=0,  \label{eqthethe}
\end{eqnarray}
\end{widetext}%
where $eq_{tt}$, $eq_{rr}$, $eq_{\varphi \varphi }$ and $eq_{zz}$ are
related to components of $tt$, $rr$, $\varphi \varphi $ and $zz$ of the
equations of motion (Eq. (\ref{FE})).

\section{Solution and Curvature Scalars}

\label{Solution_and_Curvature_Scalars}

By considering equations of motion (Eqs. (\ref{eqtt}) and (\ref{eqthethe})),
we can obtain the metric function $f\left( r\right) $ in the following form 
\begin{equation}
f\left( r\right) =-\frac{4m_{0}}{r}-\frac{\Lambda r^{2}}{3}+\frac{4q^{2}}{%
r^{2}}{+}\frac{\sigma }{r^{3\omega +1}}+\frac{g_{s}^{2}l_{s}^{2}}{r^{2}}%
\mathfrak{F}_{1},  \label{f(r)}
\end{equation}%
where $m_{0}$ is an integration constant related to geometrical mass of the
solution. Also, $\mathfrak{F}_{1}={}_{2}F_{1}\left( \left[ \frac{-1}{2},%
\frac{-1}{4}\right] ,\left[ \frac{3}{4}\right] ,-\frac{r^{4}}{l_{s}^{4}}%
\right) $ is the hypergeometric function.

In Eq.~(\ref{f(r)}), $\sigma$ sets the amplitude of the Kiselev
contribution and $\omega$ fixes its radial power. The scale $l_s$ sets
the transition between the small- and large-radius behavior of the
Letelier--Alencar term. For $r\ll l_s$, this term behaves as
$g_s^2l_s^2/r^2$ and combines with the electric contribution to give the
leading coefficient $4q^2+g_s^2l_s^2$ of $r^{-2}$ near the axis. For
$r\gg l_s$, the cloud term approaches the constant $-g_s^2$, while the
electric term decays as $r^{-2}$. These different radial profiles allow
the two contributions to affect the roots of $f(r)=0$ differently.

It is notable that the obtained solutions in Eq. (\ref{f(r)}) reduce to the
charged black string solutions when $g_{s}\rightarrow 0$, and $\sigma
\rightarrow 0$ i.e., $f\left( r\right) =-\frac{4m_{0}}{r}-\frac{\Lambda r^{2}%
}{3}+\frac{4q^{2}}{r^{2}}$.

To understand the geometrical structure, we will first check for essential
singularities by calculating the Ricci and Kretschmann scalars. The Ricci
scalar ($\mathcal{R}$) for these solutions can be expressed as follows 
\begin{equation}
\mathcal{R}=4\Lambda {+}\frac{3\omega \sigma \left( 1-3\omega \right) }{%
r^{3\left( \omega +1\right) }}{+\frac{2g_{s}^{2}\mathfrak{F}_{2}}{l_{s}^{2}}-%
}\frac{4g_{s}^{2}r^{4}\mathfrak{F}_{3}}{7l_{s}^{6}},  \label{R}
\end{equation}%

where $\mathfrak{F}_{2}={}_{2}F_{1}\left( \left[ \frac{1}{2},\frac{3}{4}%
\right] ,\left[ \frac{7}{4}\right] ,-\frac{r^{4}}{l_{s}^{4}}\right) $ and $%
\mathfrak{F}_{3}={}_{2}F_{1}\left( \left[ \frac{3}{2},\frac{7}{4}\right] ,%
\left[ \frac{11}{4}\right] ,-\frac{r^{4}}{l_{s}^{4}}\right) $. 
From Eq. (\ref{R}) we can conclude that the Ricci
scalar diverges in magnitude at the symmetry axis $(r=0)$ when $\sigma\neq0$, i.e., $\underset{r\rightarrow 0}{\lim }|\mathcal{R}|=\infty$. On the other hand, for $-1<\omega<-1/3$, the Ricci scalar tends to $4\Lambda$ at infinity.

In addition, the Kretschmann scalar ($K=\mathcal{R}_{\mu \nu \rho \sigma }%
\mathcal{R}^{\mu \nu \rho \sigma }$) of this spacetime is given by

\begin{widetext}
\begin{equation}
\begin{aligned}
K = \frac{1}{9r^{8}}\Biggl[&
  4\Bigl(12q^{2}+r\bigl(-12m_{0}-r^{3}\Lambda+3r^{-3\omega}\sigma\bigr)
      +3g_{s}^{2}l_{s}^{2}\mathfrak{F}_{1}\Bigr)^{2} \\
&+4r^{-6\omega}\Bigl(3r\sigma(1+3\omega)
    +2r^{3\omega}\bigl(12q^{2}-6m_{0}r+r^{4}\Lambda
      +3g_{s}^{2}l_{s}^{2}\mathfrak{F}_{4}\bigr)\Bigr)^{2} \\
&+r^{-6\omega}\Biggl(2r^{3\omega}\bigl(-36q^{2}+12m_{0}r+r^{4}\Lambda\bigr)
   -3r\sigma\bigl(2+9\omega(1+\omega)\bigr) \\
&\qquad\qquad
   +\frac{6g_{s}^{2}l_{s}^{2}r^{3\omega}
      \bigl(l_{s}^{4}\mathfrak{F}_{1}-2(2l_{s}^{4}+r^{4})\mathfrak{F}_{4}\bigr)}
       {l_{s}^{4}+r^{4}}\Biggr)^{2}\Biggr].
\end{aligned}
\end{equation}
\end{widetext}
where $\mathfrak{F}_{4}={}_{2}F_{1}\left( \left[ 1, \frac{5}{4}\right] ,\left[ \frac{3}{4}\right]   ,-\frac{r^{4}}{l_{s}^{4}}
\right) $. This scalar diverges at the symmetry axis $r=0$, i.e., $\underset{r\rightarrow 0}{\lim }%
\mathcal{R}_{\mu \nu \rho \sigma }\mathcal{R}^{\mu \nu \rho \sigma }\rightarrow \infty $.
\bigskip So, there is an essential curvature singularity at $r=0$. On the
other hand, for $\Lambda<0$ and $-1<\omega<-1/3$, the leading term in $f(r)$ at large $r$ is $-\Lambda r^2/3$. The curvature invariants approach their AdS values,
$\underset{r\rightarrow \infty}{\lim}\mathcal{R}=4\Lambda$ and
$\underset{r\rightarrow \infty}{\lim}\mathcal{R}_{\mu\nu\rho\sigma}\mathcal{R}^{\mu\nu\rho\sigma}=8\Lambda^2/3$. The quintessence term is subleading to $r^2$ but grows as $r^{-3\omega-1}$, so the limits of the invariants alone do not establish the standard AdS falloff conditions.

\begin{figure*}[t]
\centering
\includegraphics[width=.305\textwidth]{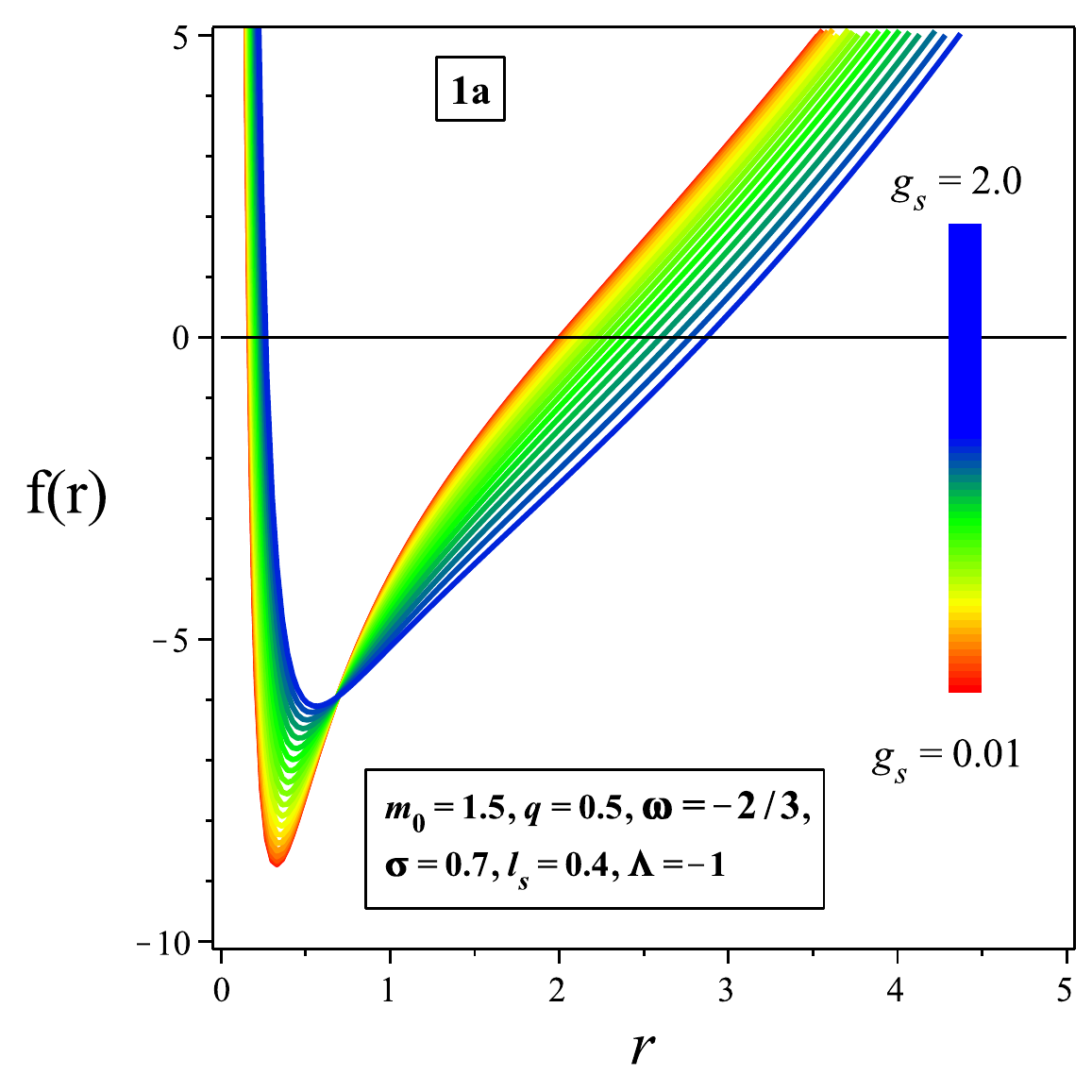} \includegraphics[width=.305\textwidth]{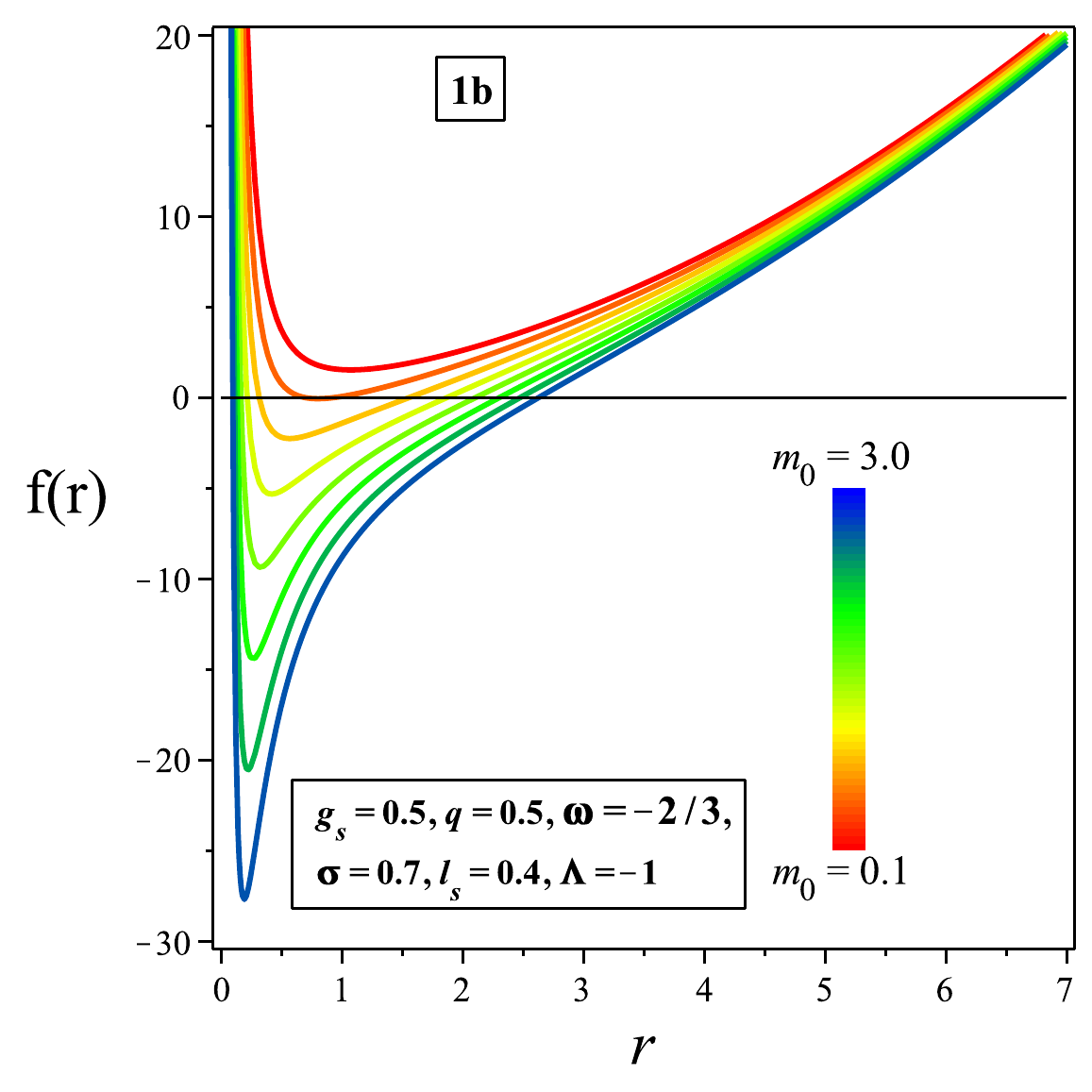} 
\includegraphics[width=.305\textwidth]{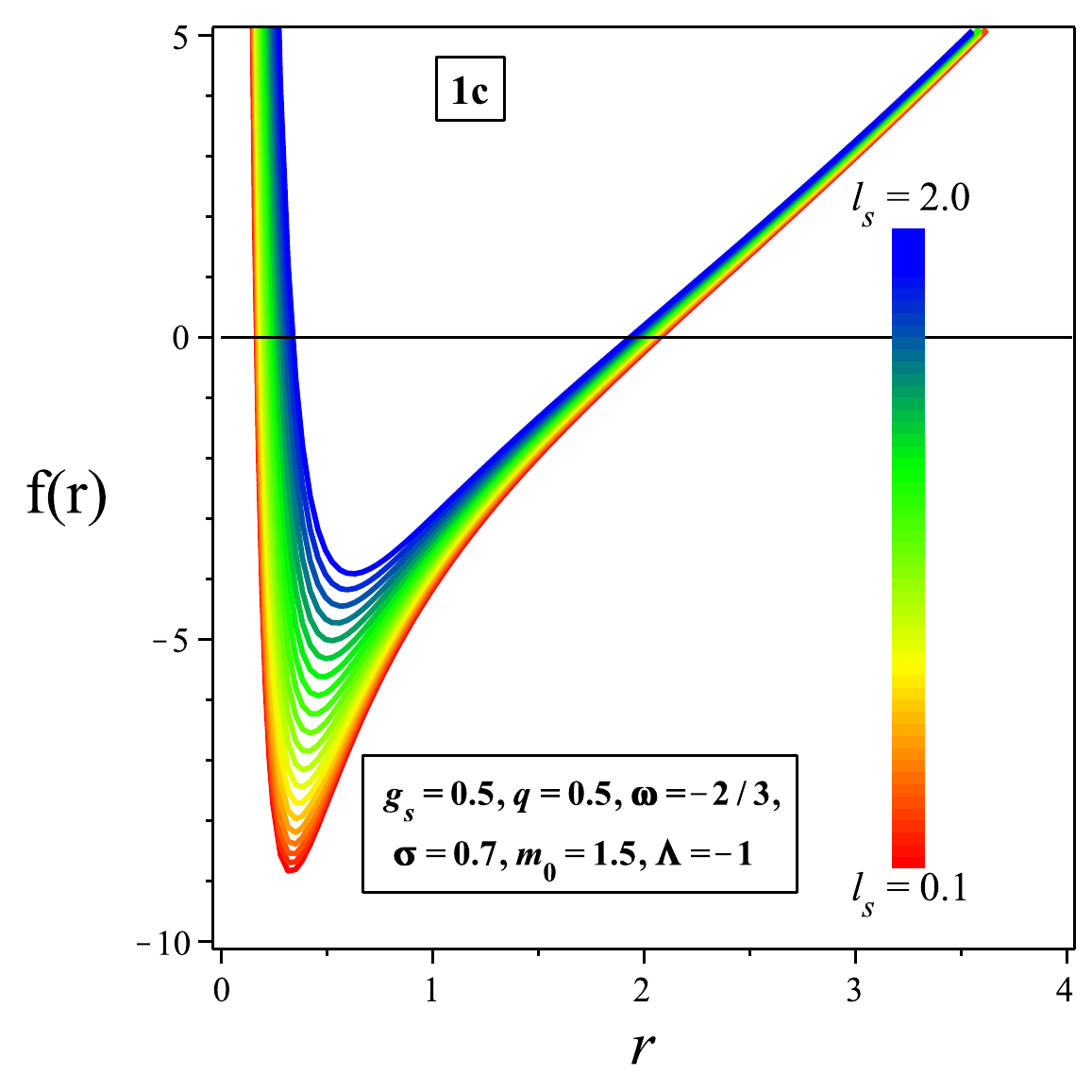} \\
\includegraphics[width=.305\textwidth]{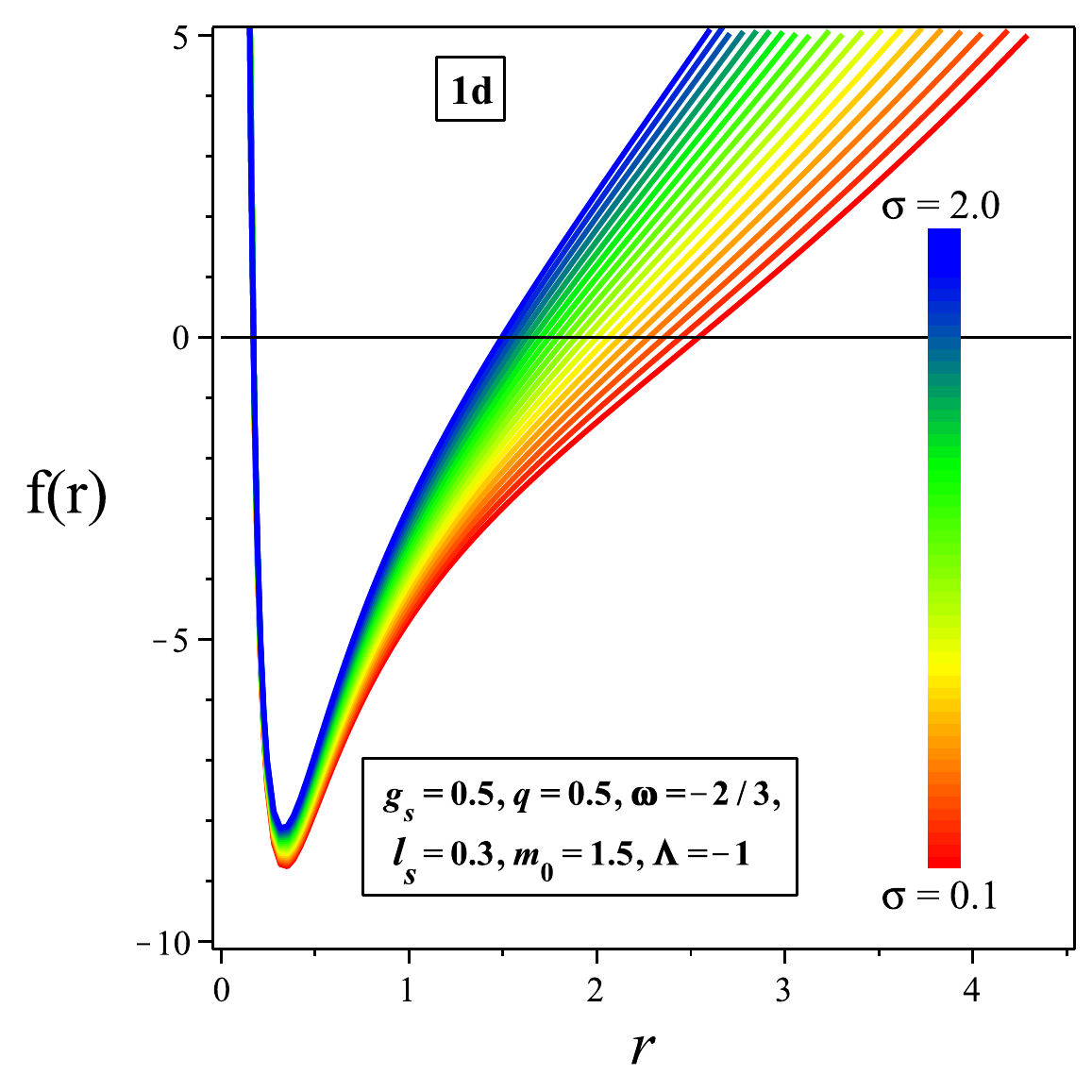} \includegraphics[width=.305\textwidth]{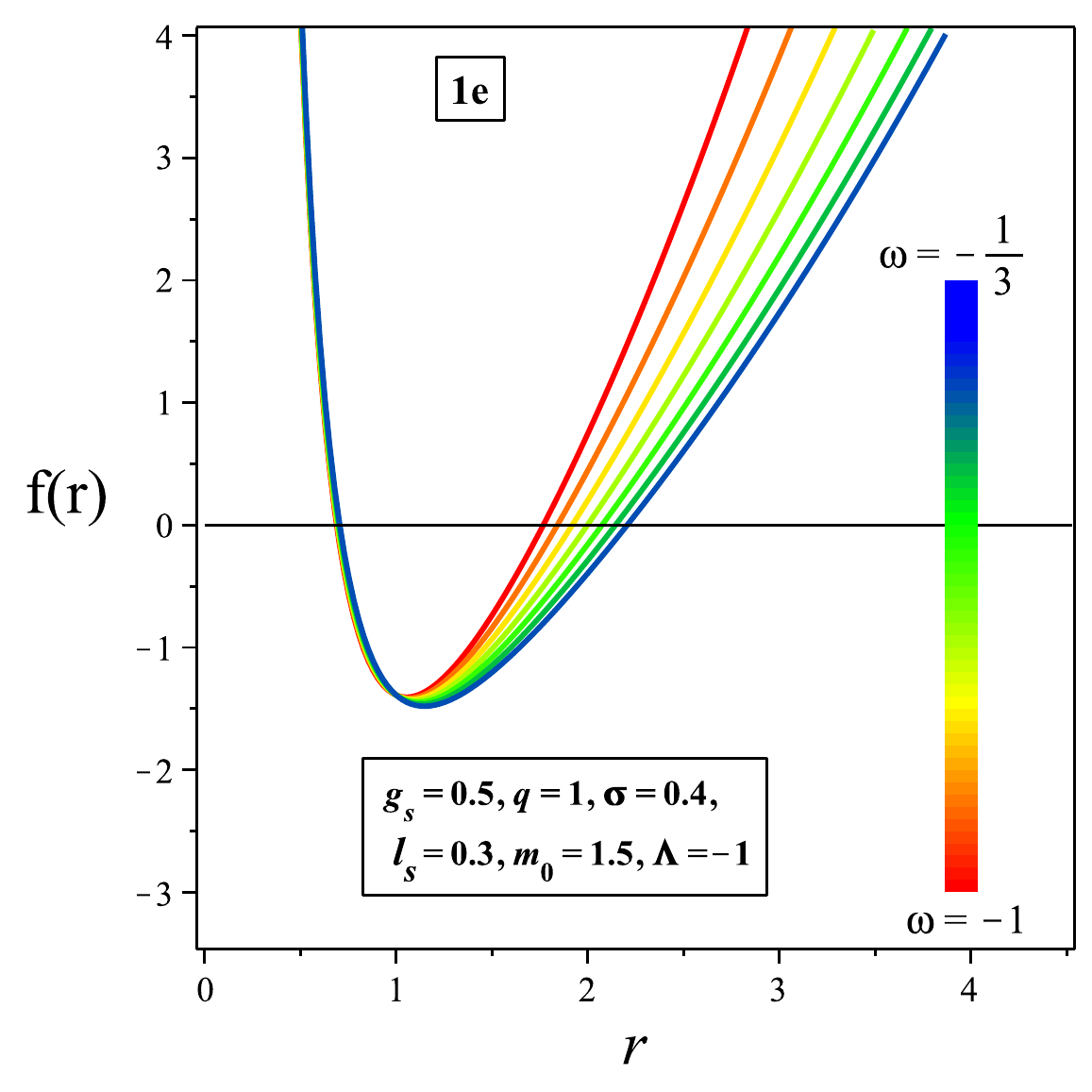} 
\includegraphics[width=.305\textwidth]{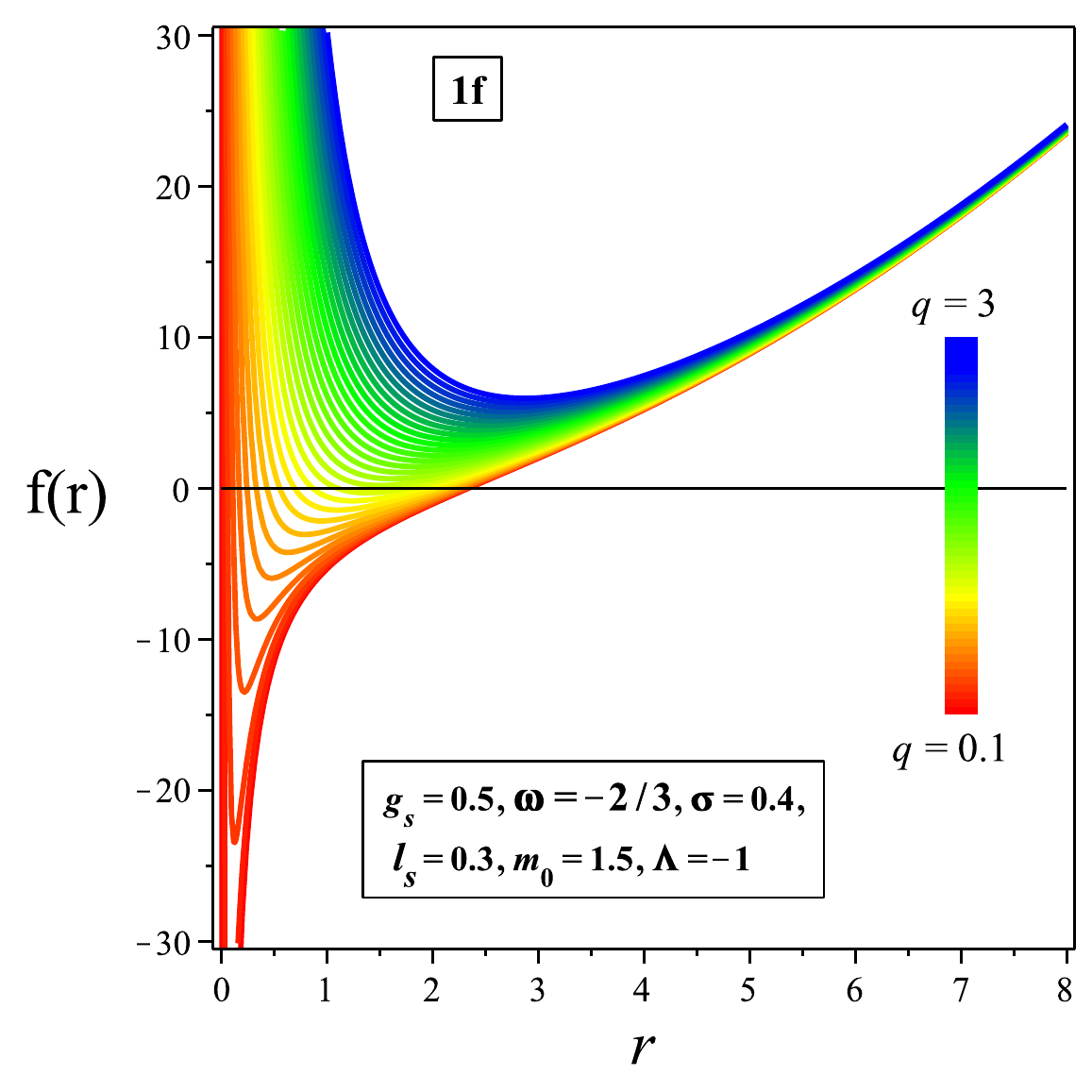}
\caption{The metric function $f(r)$ versus $r$ for different values of
parameters $g_{s}$ (\protect\ref*{fig1}a), $m_{0}$ (\protect\ref*{fig1}b), $%
l_{s}$ (\protect\ref*{fig1}c), $\protect\sigma$ (\protect\ref*{fig1}d), $%
\protect\omega$ (\protect\ref*{fig1}e), and $q$ (\protect\ref*{fig1}f).}
\label{fig1}
\end{figure*}

Our investigation of the curvature invariants indicates that the spacetime
is singular at $r=0$. Nevertheless, the metric function $f(r)$ shown in Fig. %
\ref{fig1} verifies that this axial singularity is enclosed by an event
horizon. To clarify the horizon structure, we analyze how the
Letelier--Alencar cloud of strings parameters, together with the
mass parameter, electric charge, and quintessence state parameter, modify
the zeros of $f(r)$.

The numerical behavior of the metric function shows that increasing $g_{s}$, 
$m_{0}$, and $\omega$ enlarges the event horizon radius, as can be observed
in panels \ref{fig1}a, \ref{fig1}b, and \ref{fig1}e, respectively. On the
other hand, larger values of $l_{s}$, $\sigma$, and $q$ reduce the size of
the event horizon; see panels \ref{fig1}c, \ref{fig1}d, and \ref{fig1}f. We
also find that certain model parameters possess critical values for which
the number of real roots of the metric function changes. In particular, the
geometry may contain two horizons, one degenerate horizon corresponding to
the extremal case, or no horizon at all, the latter implying a naked
singularity. Such critical behavior is exemplified in panels \ref{fig1}b and %
\ref{fig1}f.

\section{Thermodynamics}

\label{Thermodynamic}

In this section, we derive the thermodynamic characteristics of charged
Letelier--Alencar black strings in the presence of both a quintessence field
and a non-vanishing cosmological constant. Particular attention is devoted
to the effects of the Letelier--Alencar cloud of strings parameters, the
electric charge, and the cosmological constant on the thermal behavior of
the system. Our analysis further demonstrates that the obtained
thermodynamic quantities consistently satisfy the first law of black hole
thermodynamics.

Using the static Killing field $\chi^{\mu}=(1,0,0,0)$, the Hawking
temperature follows from its surface gravity $\kappa$ at the outer horizon.
\begin{equation}
\begin{aligned}
T &= \frac{\kappa}{2\pi}
   = \frac{1}{2\pi}\sqrt{\left.
        {-\frac{1}{2}}\nabla_{\mu}\chi_{\nu}\,\nabla^{\mu}\chi^{\nu}
        \right|_{r=r_{+}}}\\
  &= \left.\frac{1}{4\pi}\frac{df(r)}{dr}\right|_{r=r_{+}}.
\end{aligned}
\label{Temp1}
\end{equation}
Substituting Eq.~(\ref{f(r)}) into Eq.~(\ref{Temp1}), one arrives at the
Hawking temperature for the present class of black strings: 

\begin{equation}
\begin{aligned}
T&=-\frac{3\omega\sigma}{4\pi r_{+}^{3\omega+2}}-\frac{q^{2}}{\pi r_{+}^{3}}-%
\frac{\Lambda r_{+}}{4\pi}\\
&\quad -\frac{g_{s}^{2}}{12\pi r_{+}^{3}l_{s}^{2}}%
 \left(3l_{s}^{4}\mathfrak{F}_{1_{+}}+2r_{+}^{4}\mathfrak{F}_{2_{+}}\right),
\end{aligned}
\label{Temp2}
\end{equation}
where $\mathfrak{F}_{1_{+}}=\left.\mathfrak{F}_{1}\right|_{r=r_{+}}$ and $\mathfrak{F}_{2_{+}}=\left.\mathfrak{F}_{2}\right|_{r=r_{+}}$.
Equation (\ref%
{Temp2}) explicitly shows that the thermal properties of these black strings
are governed by the full set of metric parameters: the electric charge $q$,
the cosmological constant $\Lambda$, the Letelier--Alencar cloud of strings parameters $g_{s}$ and $l_{s}$, and the quintessence field parameters $\sigma$ and $\omega$.

We now restrict our attention to the specific case with the quintessence
equation of state parameter $\omega = -2/3$. Setting $\omega = -2/3$ in Eq.~(%
\ref{Temp2}) yields the temperature expression 

\begin{equation}
T = \frac{\sigma}{2\pi} - \frac{q^{2}}{\pi r_{+}^{3}} - \frac{\Lambda r_{+}}{%
4\pi} - \frac{g_{s}^{2}}{12\pi r_{+}^{3}l_{s}^{2}}\left(3l_{s}^{4}\mathfrak{F}_{1_{+}} + 2r_{+}^{4}\mathfrak{F}_{2_{+}}\right).
\end{equation}
In the small horizon limit ($r_{+} \to 0$), the electric charge term
dominates the temperature behavior, leading to 
\begin{equation}
\lim_{r_{+} \to 0} T \sim -\frac{q^{2}}{\pi r_{+}^{3}}-\frac{g_{s}^{2}l_{s}^{2}}{4 \pi r_{+}^{3}}\mathfrak{F}_{1_{+}},
\end{equation}
which implies that small black strings exhibit a non-physical, negative
Hawking temperature. Conversely, in the asymptotic regime of large horizons (%
$r_{+} \to \infty$), the temperature is primarily dictated by the
cosmological constant according to 
\begin{equation}
\lim_{r_{+} \to \infty} T \sim -\frac{\Lambda r_{+}}{4\pi}.
\end{equation}
Accordingly, for negative cosmological constant (i.e., $\Lambda < 0$), the
temperature remains strictly positive for large black strings.

\begin{figure*}[t]
\centering
\includegraphics[width=.30\textwidth]{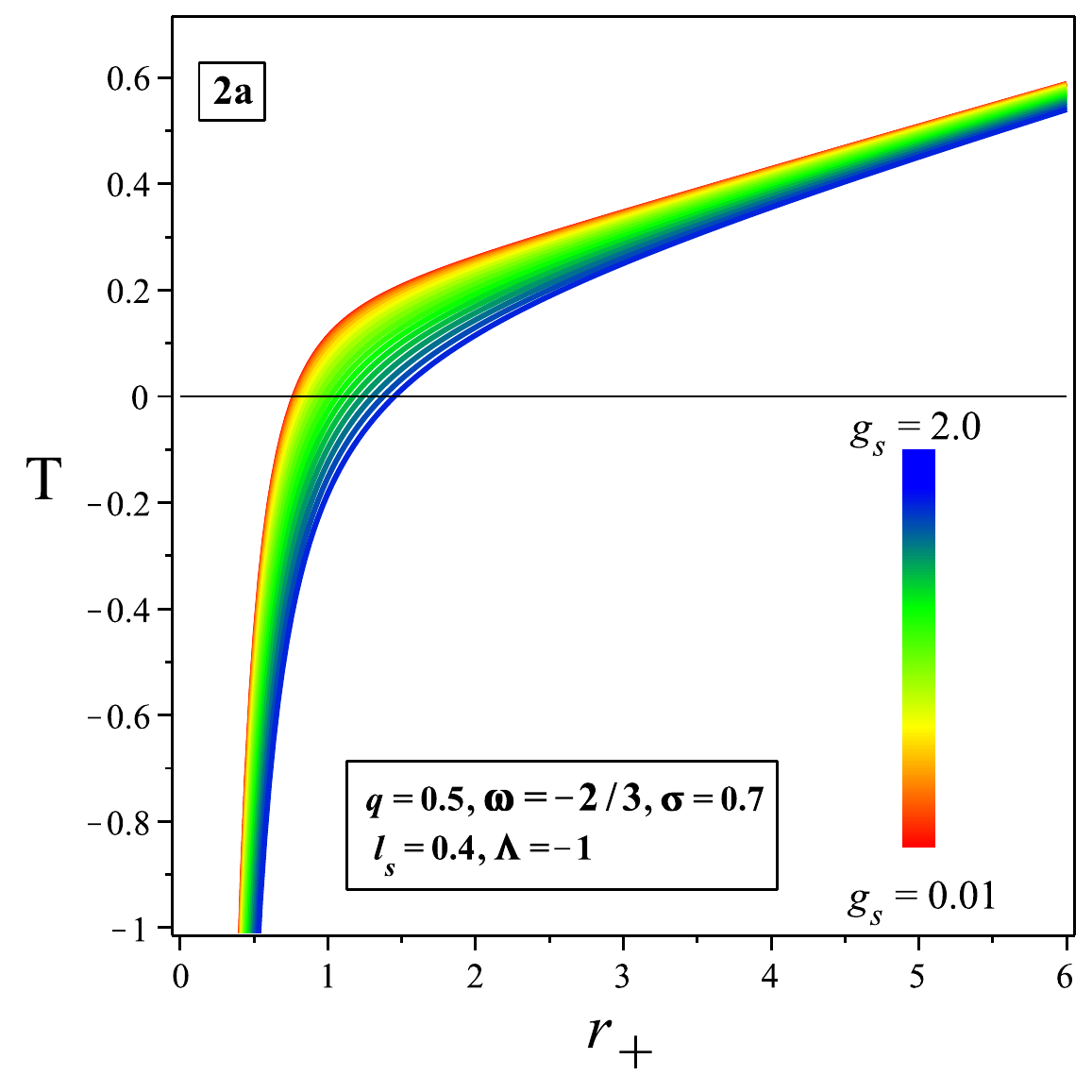} \includegraphics[width=.30\textwidth]{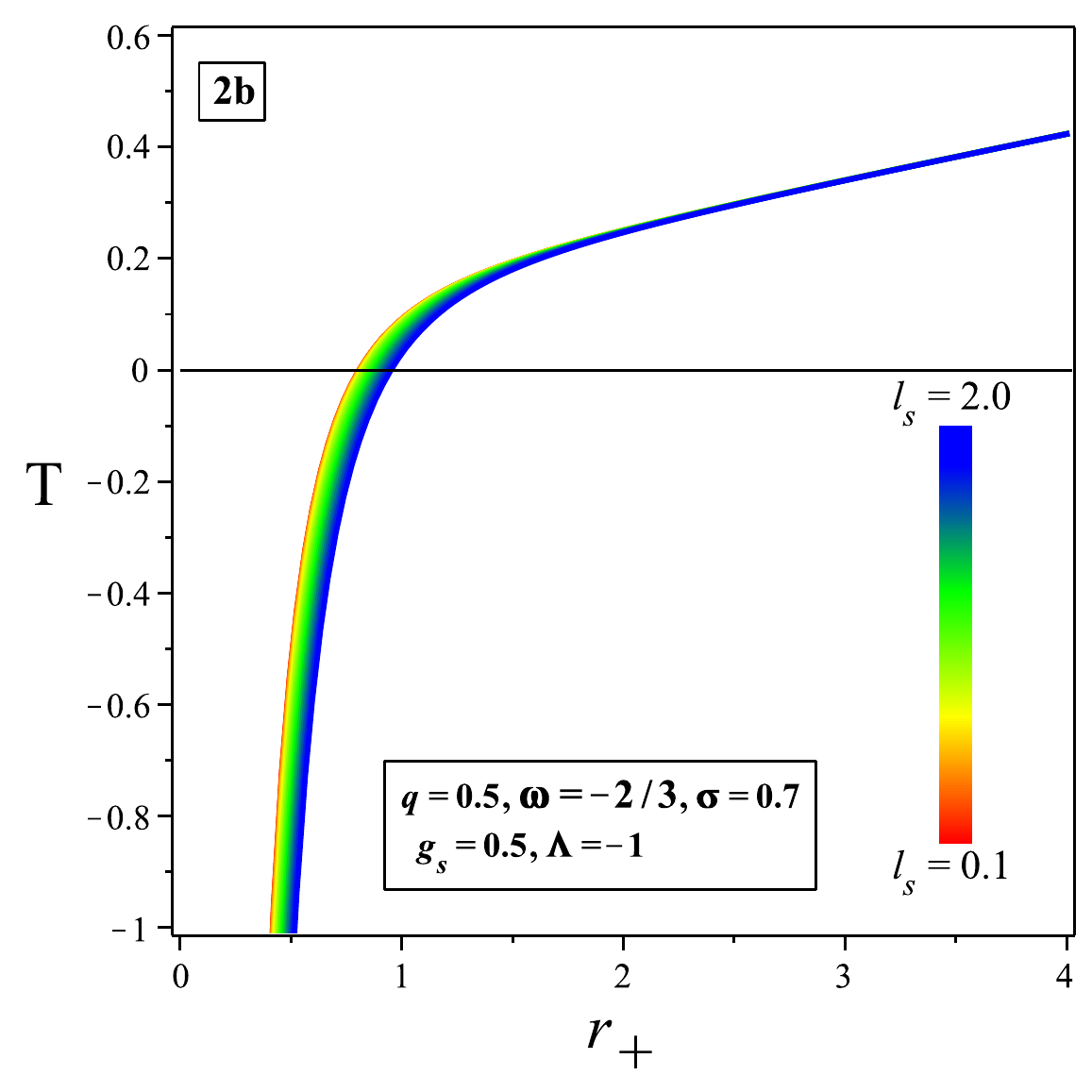} \\
\includegraphics[width=.30\textwidth]{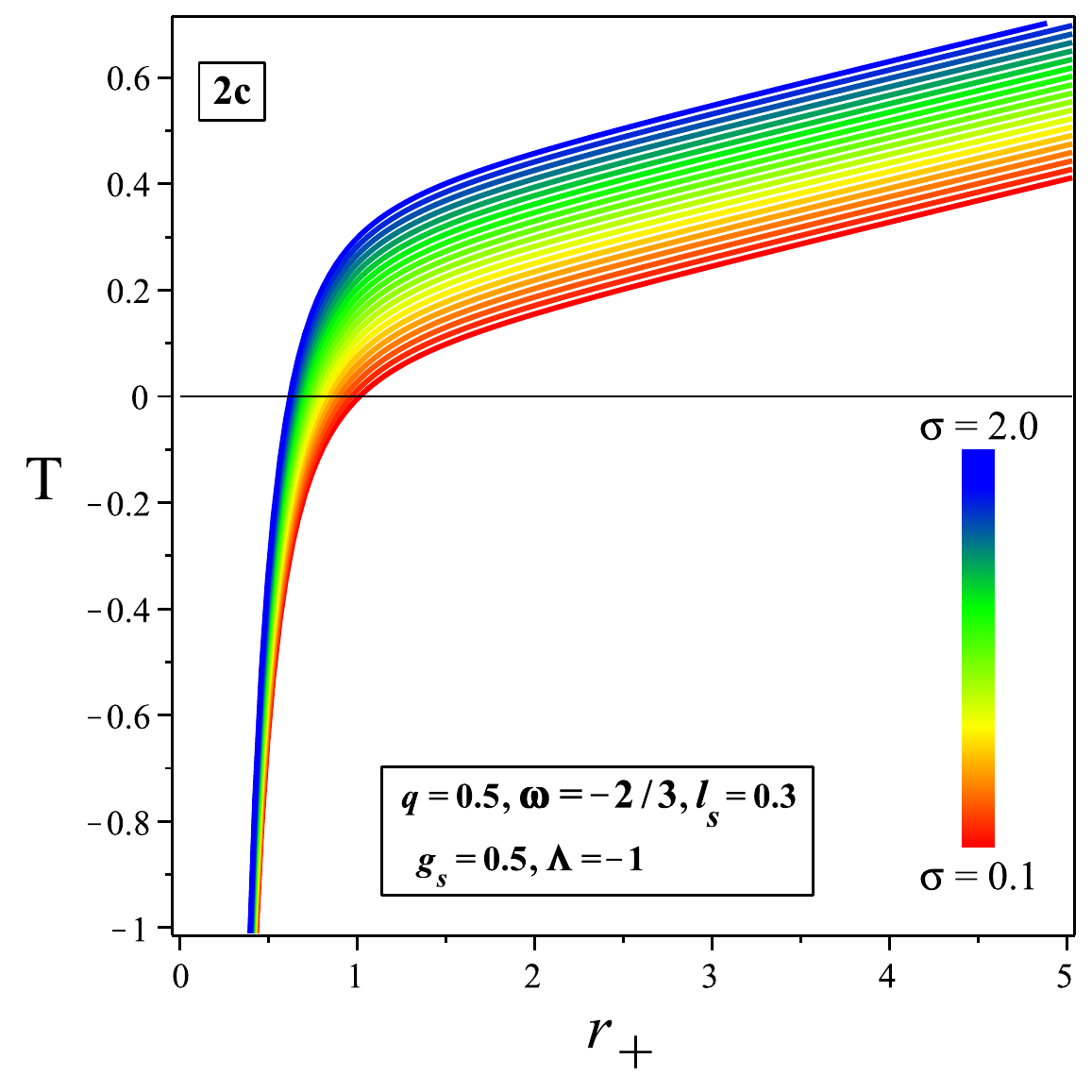} \includegraphics[width=.30\textwidth]{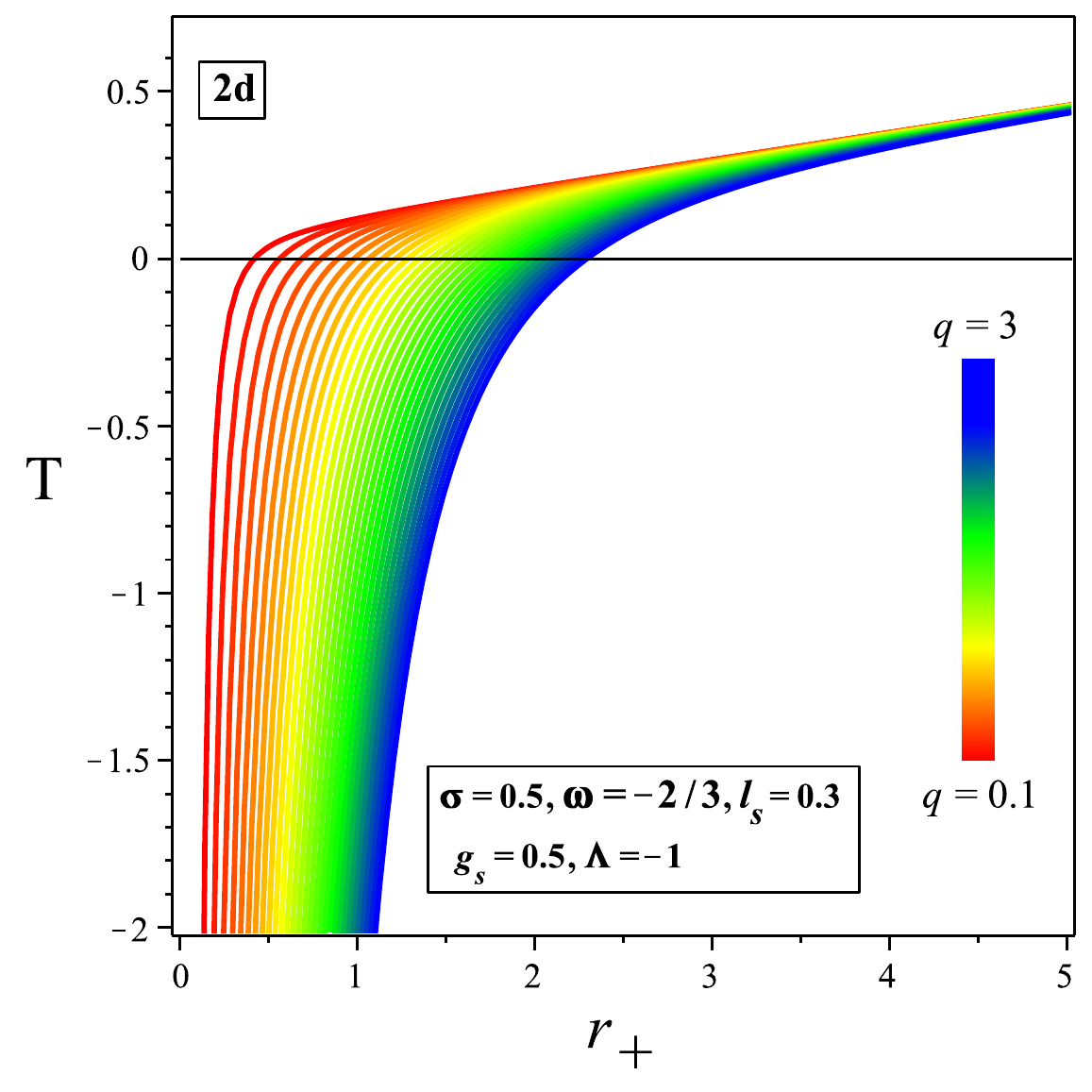}
\caption{The Hawking temperature $T$ versus $r_{+}$ for $\protect\omega=-2/3$
by considering different values of parameters $g_{s}$ (\protect\ref*{fig2}a), 
$l_{s}$ (\protect\ref*{fig2}b), $\protect\sigma$ (\protect\ref*{fig2}c), and $%
q $ (\protect\ref*{fig2}d).}
\label{fig2}
\end{figure*}

To characterize the influence of the model parameters on the thermal
behavior of the charged Letelier--Alencar black string, we display the
Hawking temperature as a function of the horizon radius in Fig.~\ref{fig2}.
The analysis reveals a single zero of the temperature, denoted by $%
r_{+_{T=0}}$, whose location depends on the choice of the physical
parameters. For horizon radii smaller than this critical value, namely $%
r_{+}<r_{+_{T=0}}$, the Hawking temperature is negative. Therefore, this
branch cannot be regarded as thermodynamically physical. In contrast, the
branch with $r_{+}>r_{+_{T=0}}$ possesses a positive temperature and
consequently represents the physically admissible black string
configurations.

Increasing the cloud of string parameters $g_{s}$ and $l_{s}$, as well as
the electric charge $q$, shifts $r_{+_{T=0}}$ toward larger values of the
horizon radius, as illustrated in panels \ref{fig2}a, \ref{fig2}b, and \ref%
{fig2}d, respectively. Hence, larger values of these parameters reduce the
interval of horizon radii for which the black string has a positive Hawking
temperature. By comparison, an increase in the quintessence normalization
parameter $\sigma $ enlarges the positive temperature domain, as can be seen
from panel \ref{fig2}c.

The entropy of the charged Letelier--Alencar black string configurations is
determined by the Bekenstein--Hawking area law, which posits a fundamental
relationship between the horizon area $A$ and the gravitational entropy 
\begin{equation}
\widetilde{S}=\frac{A}{4}.  \label{Entropy}
\end{equation}%
by evaluating the metric (\ref{metric}) at the event horizon radius $r=r_{+}$%
, the area of the cylindrical horizon is computed as follows 
\begin{equation}
A=\int_{0}^{2\pi }\int_{0}^{l_{z}}\sqrt{g_{\varphi \varphi }g_{zz}}%
\,d\varphi \,dz\bigg|_{r=r_{+}}=2\pi \alpha r_{+}^{2}l_{z}.  \label{A}
\end{equation}

To obtain a quantity independent of the longitudinal scale, we define the
entropy per unit effective string length ($\alpha l_{z}$). By combining Eqs.
(\ref{Entropy}) and (\ref{A}), the reduced entropy density $S$ is derived as 
\begin{equation}
S=\frac{\widetilde{S}}{\alpha l_{z}}=\frac{\pi r_{+}^{2}}{2}.  \label{S}
\end{equation}

We can get the electric charge of the solutions by employing the Gauss's law in
the following form 
\begin{equation}
\widetilde{Q}=\frac{F_{tr}}{4\pi }\int_{0}^{2\pi }\int_{0}^{l_{z}}\sqrt{g}%
d\varphi \,dz=q\alpha l_{z},  \label{Q}
\end{equation}%
where for the case $t=$ constant and $r=$constant, the determinant of metric
tensor $g$ is $\alpha ^{2}r^{4}$. Also, the Faraday tensor is $F_{tr}=\frac{%
2q}{r^{2}}$. We obtain the electric charge per unit effective string length (%
$\alpha l_{z}$), in the following form 
\begin{equation}
Q=\frac{\widetilde{Q}}{\alpha l_{z}}=q.  \label{QQ}
\end{equation}

In order to obtain electric potential, we can calculate it on the horizon
with respect to a reference which leads to 
\begin{equation}
\Phi =A_{\mu }\chi ^{\mu }\left\vert _{r\rightarrow \infty }\right. -A_{\mu
}\chi ^{\mu }\left\vert _{r\rightarrow r_{+}}\right. =\frac{2q}{r_{+}},
\label{U}
\end{equation}%
where the gauge potential is obtained as $A_{\mu }=\left( h\left( r\right)
,0,0,0\right) =\left( \frac{-2q}{r},0,0,0\right) $.

Applying Ashtekar-Magnon-Das (AMD) approach, we find the total mass of these
black strings in the following form%
\begin{equation}
M=m_{0}=\frac{q^{2}}{r_{+}}-\frac{\Lambda r_{+}^{3}}{12}+\frac{\sigma }{%
4r_{+}^{3\omega }}+\frac{g_{s}^{2}l_{s}^{2}}{4r_{+}}\mathfrak{F}_{1_{+}},
\label{MM}
\end{equation}%
where $m_{0}$ is the geometrical mass and is given by solving $\left.
f(r)\right\vert _{r=r_{+}}=0$.

To examine the mass profile of the black string, we set $\omega=-2/3$ in Eq.~(\ref{MM}), which gives
\begin{equation}
M=\frac{q^{2}}{r_{+}}-\frac{\Lambda r_{+}^{3}}{12}+\frac{\sigma r_{+}^{2}}{4}+\frac{g_{s}^{2}l_{s}^{2}}{4r_{+}}\mathfrak{F}_{1_{+}},
\end{equation}
and for sufficiently small horizon radius, the electric charge term dominates, so that
\begin{equation}
\lim_{r_{+}\rightarrow 0} M \sim \frac{q^{2}}{r_{+}}+\frac{g_{s}^{2}l_{s}^{2}}{4r_{+}}\mathfrak{F}_{1_{+}},
\end{equation}
which is positive. Hence, the small black string configurations possess positive mass. In the opposite limit, when $r_{+}\rightarrow \infty$, the leading contribution is controlled by the cosmological constant and behaves as
\begin{equation}
\lim_{r_{+}\rightarrow \infty} M \sim -\frac{\Lambda r_{+}^{3}}{12},
\end{equation}
so, for the negative cosmological constant ($\Lambda<0$), the mass remains positive at large horizon radii. The structure of Eq.~(\ref{MM}) also implies the existence of a minimum mass in the intermediate regime, and this extremal value depends on the parameters $g_{s}$, $l_{s}$, $q$, and $\sigma$.

Replacing Eqs. (\ref{S}) and (\ref{QQ}) within Eq. (\ref{MM}), we can
obtain $M=M(S,Q)$ in the following form 
\begin{equation}
M(S,Q)=\frac{Q^{2}}{\sqrt{\frac{2S}{\pi }}}-\frac{\Lambda S^{3/2}}{3\sqrt{2}%
\pi ^{3/2}}+\frac{\sigma }{4\left( \frac{2S}{\pi }\right) ^{3\omega /2}}+%
\frac{g_{s}^{2}l_{s}^{2}}{4\sqrt{2}\sqrt{\frac{S}{\pi }}}\mathfrak{F}%
_{1_{S}},  \label{MS}
\end{equation}%
where $\mathfrak{F}_{1_{S}}=\left. \mathfrak{F}_{1_{+}}\right\vert _{r_{+}=%
\sqrt{\frac{2S}{\pi }}}=_{2}F_{1}\left( \left[ \frac{-1}{2},\frac{-1}{4}%
\right] ,\left[ \frac{3}{4}\right] ,-\frac{4S^{2}}{\pi ^{2}l_{s}^{4}}\right)$. Calculating derivatives of mass ($M$) with respect to entropy ($S$) and
charge ($Q$) yields temperature ($T$) and electric potential ($\Phi $),
respectively. Although the parameters of Letelier--Alencar and charge
affected thermodynamic and conserved quantities, the first law remains valid
as 
\begin{equation}
dM=TdS+\Phi dQ.
\end{equation}

\subsection{Local Thermodynamic Stability}

Within the framework of the canonical ensemble, the local thermodynamic stability of a system, such as a black string, is fundamentally dictated by the sign of its heat capacity. A positive heat capacity signifies a locally stable state, whereas a negative value indicates instability. Consequently, we perform a detailed analysis of the heat capacity to assess the stability criteria for charged Letelier--Alencar black strings, accounting for the contributions from the quintessence field and the cosmological constant.

The heat capacity is defined as $C=\frac{T}{\left( \frac{\partial T}{\partial S}\right) _{q}}=T\left( \frac{\frac{\partial S}{\partial r_{+}}}{\frac{\partial T}{\partial r_{+}}}\right) _{q}$. By applying Eqs. (\ref{Temp2}), and (\ref{S}) within $C$, we can get the heat capacity as 
\begin{widetext}
\begin{equation}
 C=\frac{7\pi \left( 3\Lambda
 l_{s}^{2}r_{+}^{4}+12q^{2}l_{s}^{2}+3g_{s}^{2}l_{s}^{4}\mathfrak{F}%
 _{1_{+}}+2g_{s}^{2}r_{+}^{4}\mathfrak{F}_{2_{+}}+\frac{9\sigma \omega
 l_{s}^{2}}{r_{+}^{3\omega -1}}\right) }{3l_{s}^{2}\left( 7\Lambda r_{+}^{2}-\frac{84q^{2}}{r_{+}^{2}}-\frac{21g_{s}^{2}l_{s}^{2}}{r_{+}^{2}}\mathfrak{F} _{1_{+}}-\frac{4g_{s}^{2}r_{+}^{6}}{l_{s}^{6}}\mathfrak{F}_{3_{+}}-\frac{ 21\omega \left( 3\omega +2\right) \sigma }{r_{+}^{3\omega +1}}\right) }.
\label{C}
\end{equation}
\end{widetext}

We now address the effects of the Letelier--Alencar cloud of strings, the electric charge, and the quintessence field on the local thermodynamic stability of the black string. The calculations show that the heat capacity possesses a single zero, denoted as $r_{+, C=0}$. For horizon radii smaller than this threshold ($r_{+} < r_{+, C=0}$), the heat capacity is negative, which indicates that small black strings are thermodynamically unstable. Conversely, for larger configurations satisfying $r_{+} > r_{+, C=0}$, the heat capacity becomes positive, indicating that large black strings satisfy the local thermodynamic stability criterion.

\begin{figure*}[t]
\centering
\includegraphics[width=.30\textwidth]{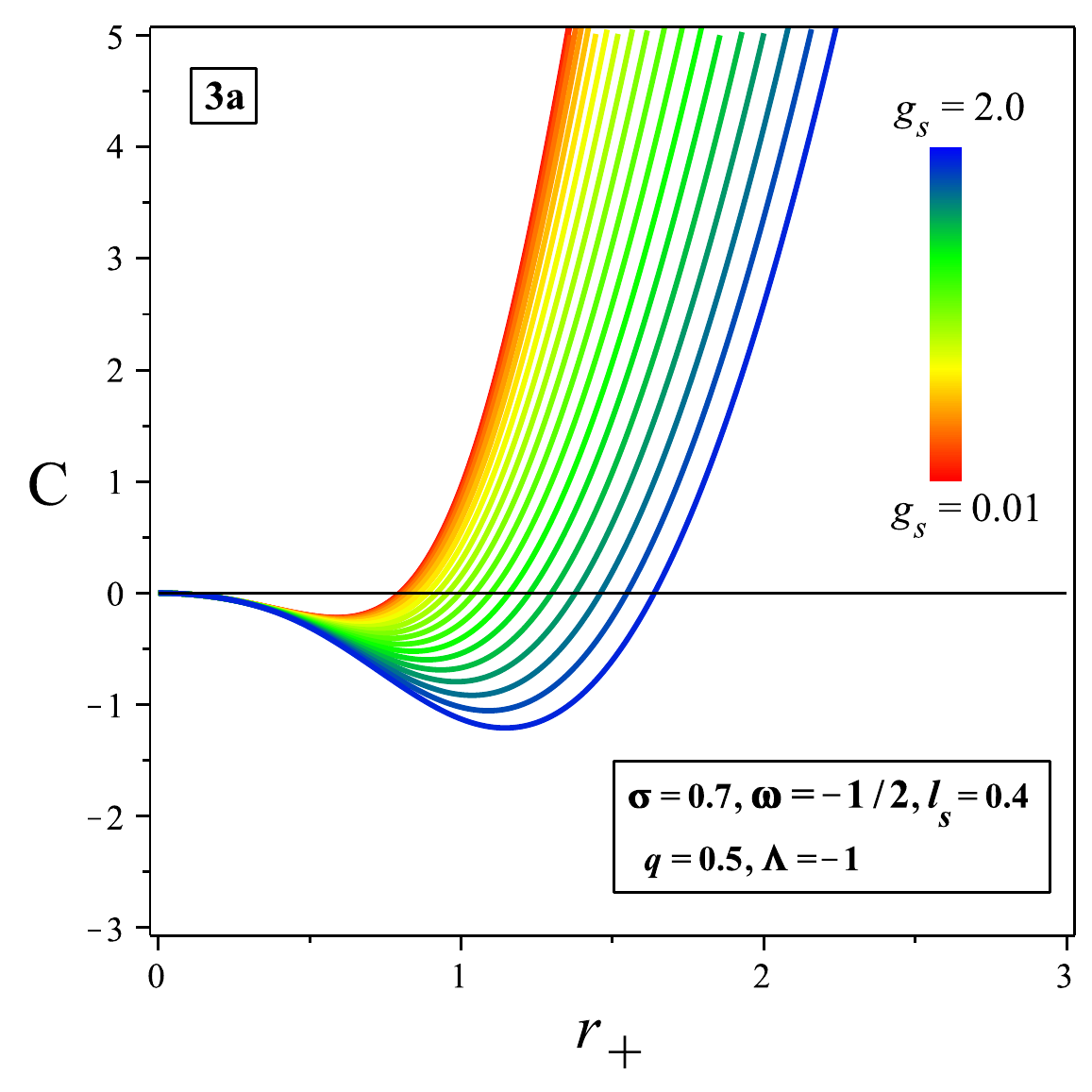} \includegraphics[width=.30\textwidth]{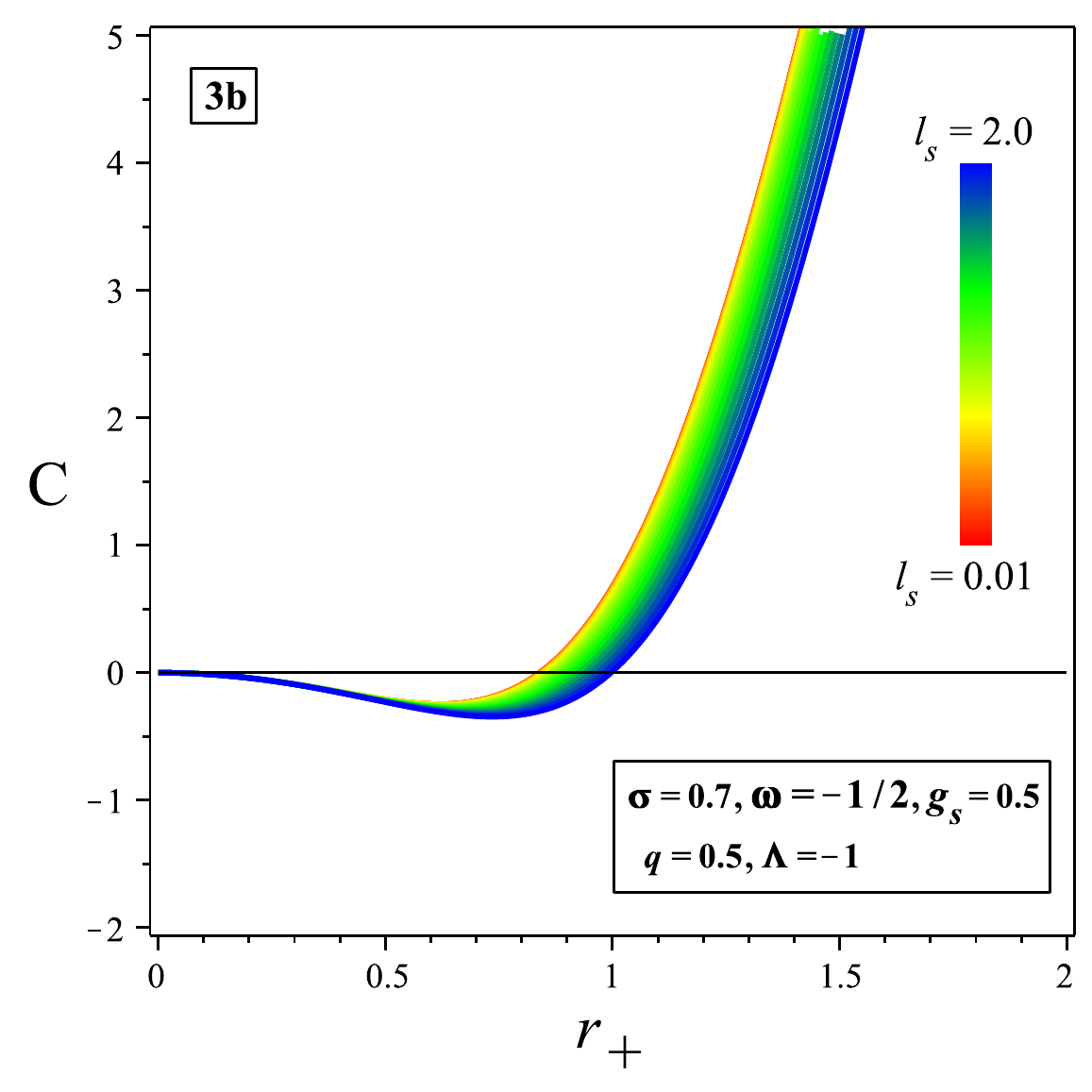} \\
\includegraphics[width=.30\textwidth]{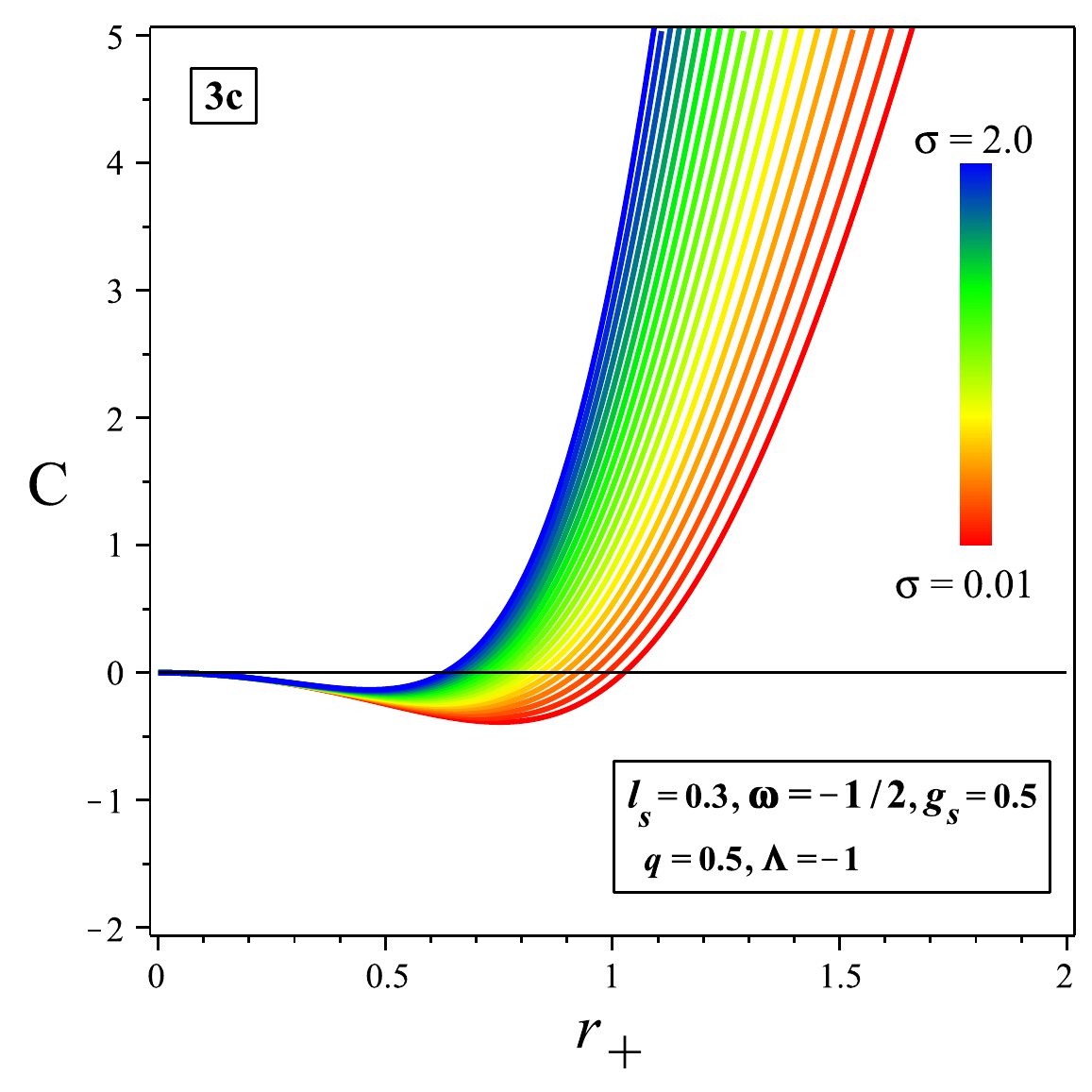} \includegraphics[width=.30\textwidth]{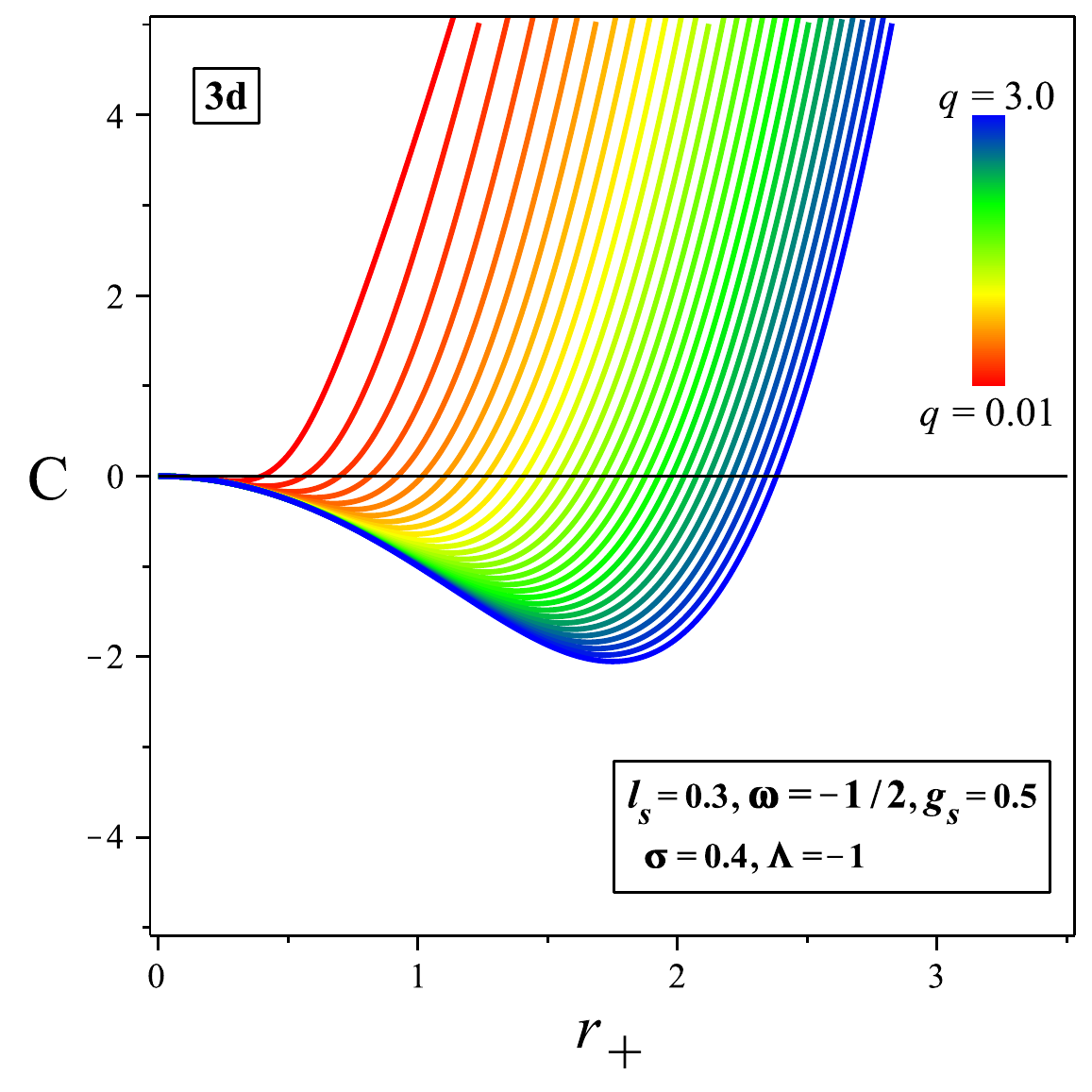}
\caption{The heat capacity $C$ versus $r_{+}$ for $\protect\omega=-1/2$ by considering different values of parameters $g_{s}$ (\protect\ref*{fig3}a), $l_{s}$ (\protect\ref*{fig3}b), $\protect\sigma$ (\protect\ref*{fig3}c), and $q$ (\protect\ref*{fig3}d).}
\label{fig3}
\end{figure*}

Figure \ref{fig3} illustrates how the different model parameters modify the domain of local stability. Our analysis shows that these parameters influence the zero of the heat capacity in two distinct ways
\begin{enumerate}
	\item[i)] The region of local stability shrinks as $g_{s}$, $l_{s}$, and $q$ increase. Specifically, the zero of the heat capacity shifts toward larger radii with increasing values of $g_{s}$, $l_{s}$, and $q$ (see panels \ref{fig3}a, \ref{fig3}b, and \ref{fig3}d, respectively).
	\item[ii)] The domain of local stability expands when the quintessence normalization parameter $\sigma$ increases. In this case, the zero of the heat capacity shifts toward smaller radii, as shown in panel \ref{fig3}c.
\end{enumerate}

\subsection{Global Thermodynamic Stability}

In the context of the grand-canonical ensemble, the global stability of a
thermodynamic system can be studied by Gibbs's potential. In other words,
the negative of the Gibbs potential determines the global stability of a
thermodynamic system. The Gibbs potential is defined as 
\begin{equation}
G=M-TS-\Phi Q.  \label{G}
\end{equation}

Using Eqs. (\ref{Temp2}), (\ref{S}), (\ref{QQ}), (\ref{U}), and (\ref%
{MM}) within Eq. (\ref{G}), we find the Gibbs potential in the following form 
\begin{equation*}
\begin{aligned}
G&=\frac{\left( 3\omega +2\right) \sigma }{8r_{+}^{3\omega }}\\
&\quad+\frac{\Lambda
	l_{s}^{2}r_{+}^{4}-12Q^{2}l_{s}^{2}+9g_{s}^{2}l_{s}^{4}\mathfrak{F}%
	_{1_{+}}+2g_{s}^{2}r_{+}^{4}\mathfrak{F}_{2_{+}}}{24l_{s}^{2}r_{+}},
\end{aligned}
\end{equation*}
This expression depends on various parameters such as $g_{s}$, $l_{s}$, $\sigma$, $\omega$ and $Q$.

\begin{figure*}[t]
\centering
\includegraphics[width=.30\textwidth]{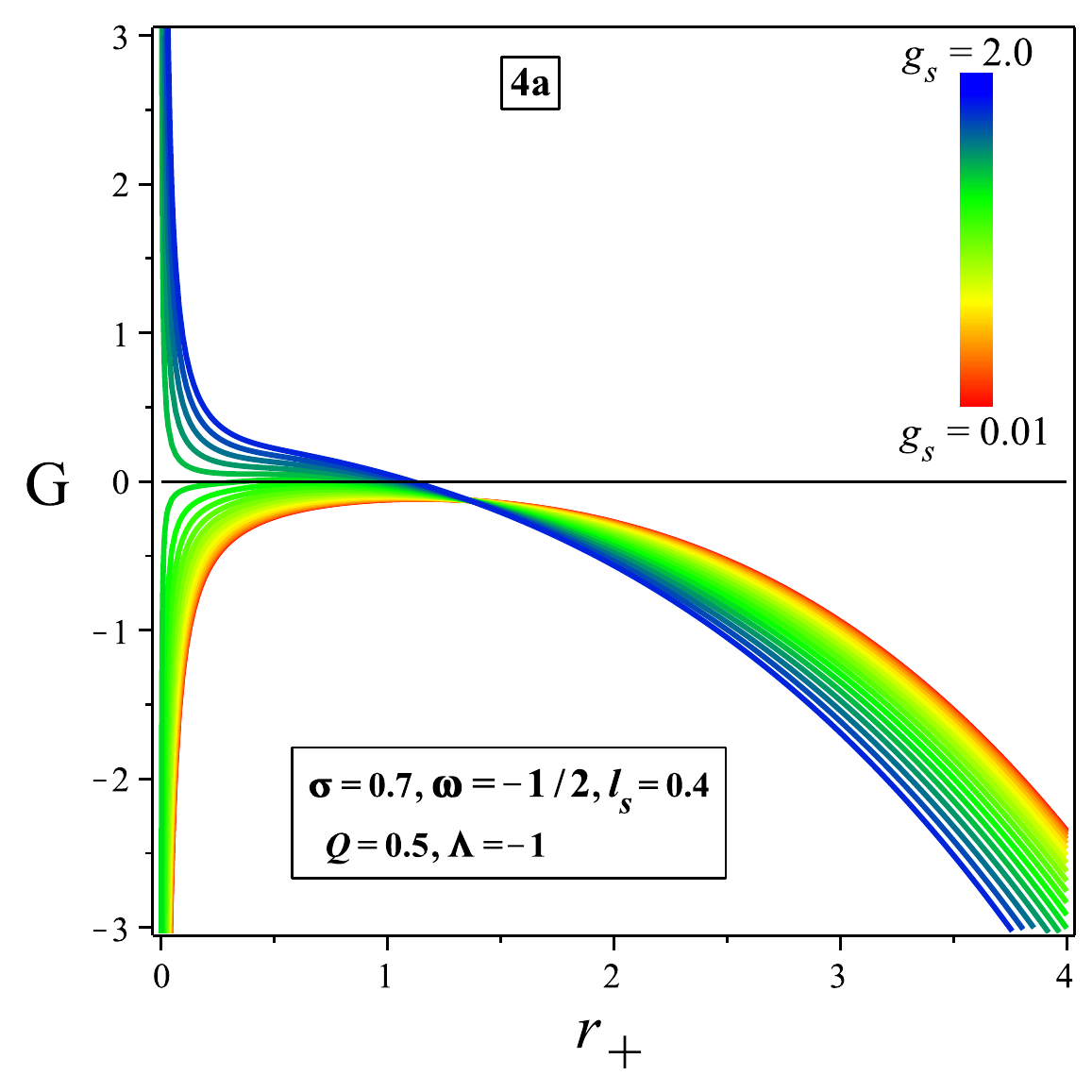} \includegraphics[width=.30\textwidth]{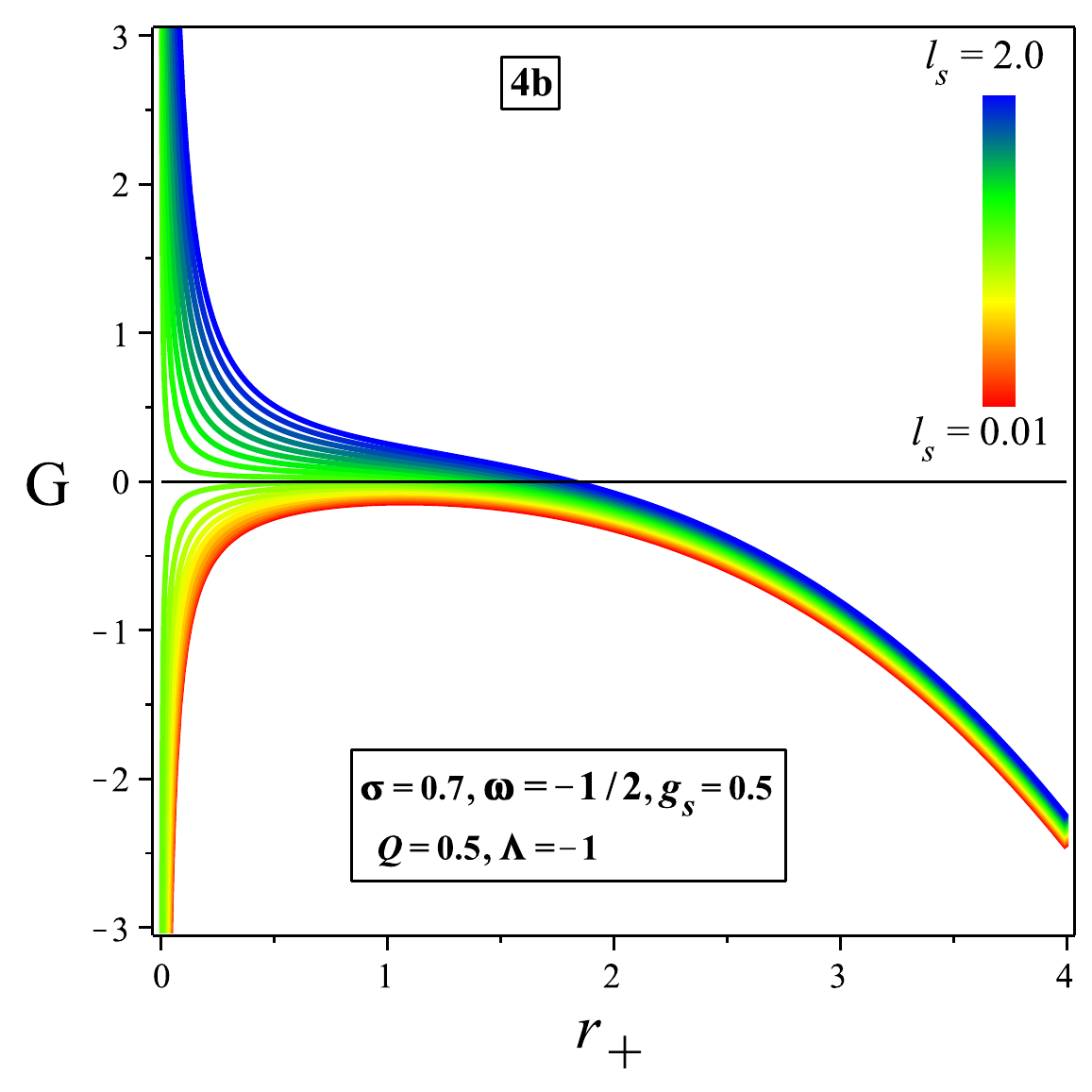} \\
\includegraphics[width=.30\textwidth]{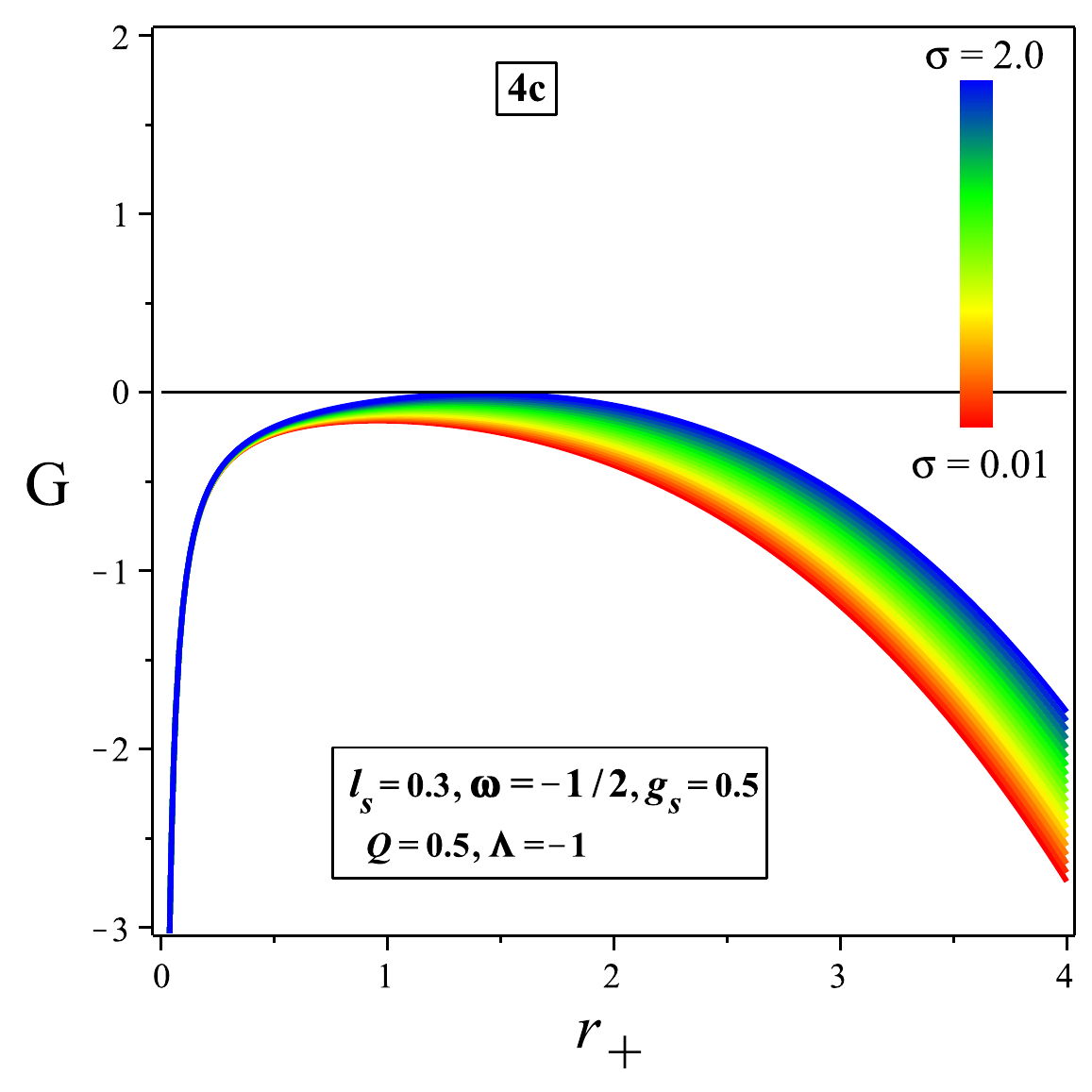} \includegraphics[width=.30\textwidth]{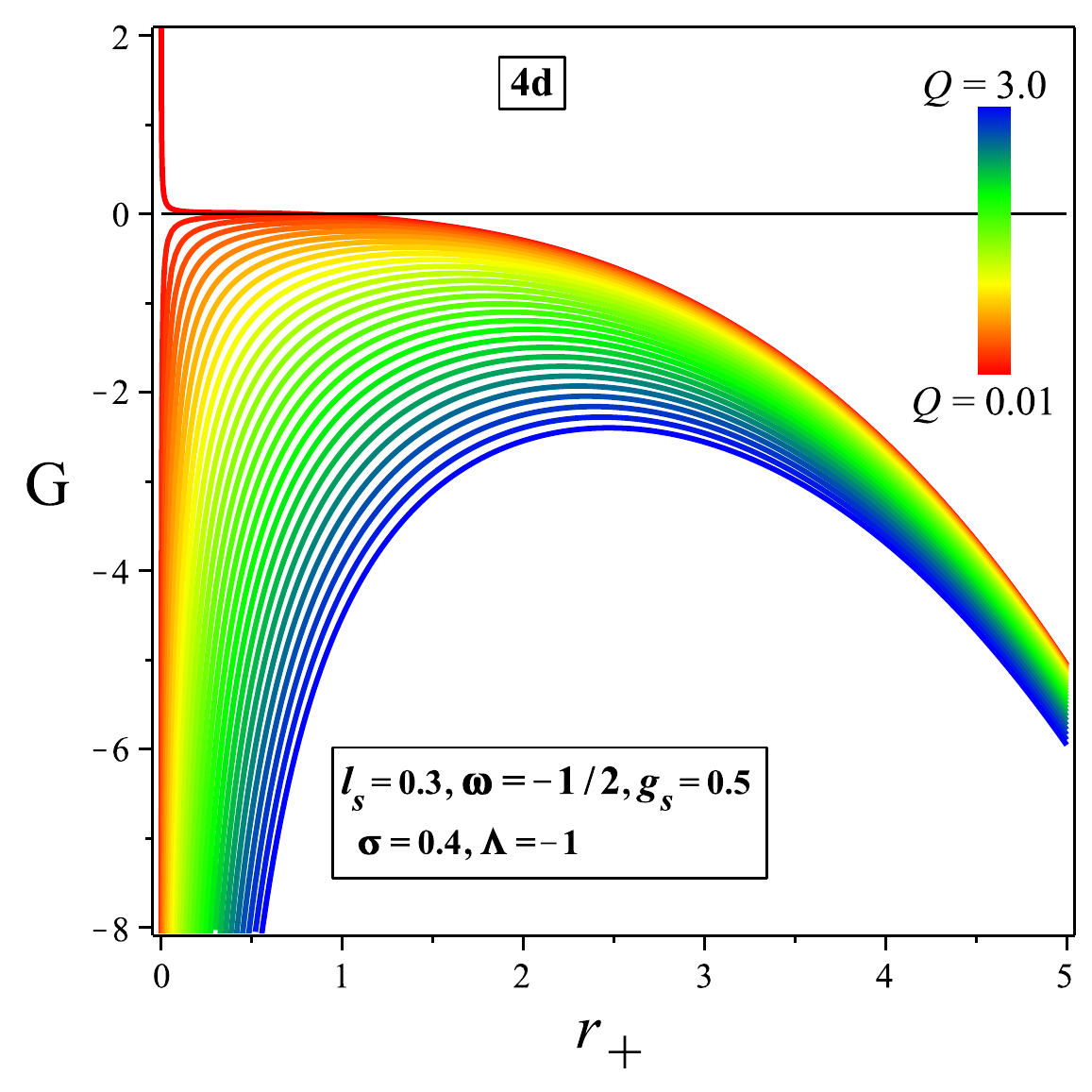}
\caption{The Gibbs potential $G$ versus $r_{+}$ for $\protect\omega=-1/2$ by considering different values of parameters $g_{s}$ (\protect\ref*{fig4}a), $l_{s}$ (\protect\ref*{fig4}b), $\protect\sigma$ (\protect\ref*{fig4}c), and $Q$ (\protect\ref*{fig4}d).}
\label{fig4}
\end{figure*}

We now examine how the model parameters affect the Gibbs potential, as displayed in Fig.~\ref{fig4}. The behavior of the Gibbs potential shows that the global thermodynamic stability of the black string depends sensitively on the parameters of the system. Our analysis can be summarized as follows:
\begin{enumerate}
	\item[i)] \textbf{Role of $g_{s}$:} There exists a critical value of the parameter cloud of strings, denoted by $g_{s_{c}}$. For $g_{s}<g_{s_{c}}$, the Gibbs potential has no real root and remains negative over the whole range of the horizon radius, implying that the corresponding black string configurations are globally stable. Once $g_{s}>g_{s_{c}}$, however, a single zero of the Gibbs potential appears. In this case, $G$ is positive for radii smaller than the root and negative for larger radii. Therefore, sufficiently small black strings fail to satisfy the global stability condition, whereas large black strings remain globally stable (see panel \ref{fig4}a).
	
	\item[ii)] \textbf{Role of $l_{s}$:} The nonlinear parameter $l_{s}$ exhibits a qualitatively similar influence. There is a critical value $l_{s_{c}}$ such that for $l_{s}<l_{s_{c}}$, the Gibbs potential stays negative everywhere, indicating global stability throughout the allowed range. By contrast, when $l_{s}>l_{s_{c}}$, the Gibbs potential develops one root. In this regime, $G>0$ before the root and $G<0$ beyond it. As a result, small black strings are globally unstable, while large black strings satisfy the condition for global stability (see panel \ref{fig4}b).
	
	\item[iii)] \textbf{Role of $\sigma$:} For the values of the quintessence normalization parameter $\sigma$ considered here, the Gibbs potential remains negative for all horizon radii. This behavior indicates that the black string solutions preserve global thermodynamic stability over the entire domain (see panel \ref{fig4}c).
	
	\item[iv)] \textbf{Role of $q$:} The electric charge also admits a critical value, denoted by $q_{c}$. For $q<q_{c}$, the Gibbs potential possesses a single root: it is positive for smaller radii and becomes negative after crossing the zero. Hence, very small black strings are globally unstable, whereas sufficiently large black strings satisfy the condition $G<0$ and are globally stable. For $q>q_{c}$, the Gibbs potential is negative everywhere, showing that highly charged black strings are globally stable across the full range of horizon radii (see panel~\ref{fig4}d).
\end{enumerate}

\subsection{Thermodynamics in Extended Phase Space}

Under the extended phase space formalism, the cosmological constant is
recast as a thermodynamic variable, directly analogous to pressure ($P=\frac{%
-\Lambda }{8\pi }$) \cite{PLambda2,PLambda3,PLambda4}. This reinterpretation
consequently posits black hole/string mass as a manifestation of enthalpy.

Following the substitution of $\Lambda =-8\pi P$ into Equation (\ref{MS}),
the subsequent analysis reveals 
\begin{equation}
\begin{aligned}
&\mathcal{M}=M\left( S,Q,P,\sigma ,\omega ,g_{s},l_{s}\right)\\
&=\frac{Q^{2}}{%
\sqrt{\frac{2S}{\pi }}}+\frac{4\sqrt{2}S^{3/2}P}{3\sqrt{\pi }}\\
&\quad+\frac{\sigma 
}{4\left( \frac{2S}{\pi }\right) ^{3\omega /2}}+\frac{g_{s}^{2}l_{s}^{2}}{4%
\sqrt{2}\sqrt{\frac{S}{\pi }}}\mathfrak{F}_{1_{S}}.
\end{aligned}  \label{MSQPb}
\end{equation}

From Eq. (\ref{MSQPb}), we derive the conjugate quantities corresponding to
the intensive parameters $S$, $Q$, $P$, $g_{s}$, and $l_{s}$. These are: 
{\allowdisplaybreaks[1]
\begin{align}
T &=\left( \frac{\partial \mathcal{M}}{\partial S}\right)
_{Q,P,g_{s},l_{s}}\notag\\
&=2\sqrt{2}P\sqrt{\frac{S}{\pi }}-\frac{\sqrt{\pi }Q^{2}}{2%
	\sqrt{2}S^{3/2}}\notag\\
&\quad-\frac{3\omega \sigma }{8S\left( \frac{2S}{\pi }\right)
	^{3\omega /2}}\notag\\
&\quad-\frac{g_{s}^{2}\sqrt{S}}{3\sqrt{2}l_{s}^{2}\pi ^{3/2}}\left( 
\frac{3\pi ^{2}l_{s}^{4}}{8S^{2}}\mathfrak{F}_{1_{S}}+\mathfrak{F}%
_{2_{S}}\right) ,  \label{TSQPb} \\
V &=\left( \frac{\partial \mathcal{M}}{\partial P}\right)
_{S,Q,g_{s},l_{s}}=\frac{4S}{3}\sqrt{\frac{2S}{\pi }},  \label{VSQPb} \\
\Phi &=\left( \frac{\partial \mathcal{M}}{\partial Q}\right)
_{S,P,g_{s},l_{s}}=\frac{Q}{\sqrt{\frac{S}{2\pi }}},  \label{PhiSQPb} \\
\mathcal{G} &=\left( \frac{\partial \mathcal{M}}{\partial g_{s}}\right)
_{S,Q,P,l_{s}}=\sqrt{\frac{\pi }{8S}}g_{s}l_{s}^{2}\mathfrak{F}_{1_{S}},
\label{GSQPb} \\
\mathcal{L} &=\left( \frac{\partial \mathcal{M}}{\partial l_{s}}\right)
_{S,Q,P,g_{s}}\notag\\
&=\frac{\sqrt{2}g_{s}^{2}S^{3/2}}{3l_{s}^{3}\pi ^{3/2}}\left( 
\frac{3\pi ^{2}l_{s}^{4}}{4S^{2}}\mathfrak{F}_{1_{S}}+\mathfrak{F}%
_{2_{S}}\right) ,  \label{LSQPb} \\
\Sigma &=\left( \frac{\partial \mathcal{M}}{\partial \sigma }\right)
_{S,Q,P,g_{s}}=\frac{1}{4\left( \frac{2S}{\pi }\right) ^{3\omega /2}},
\label{SigSQPb} \\
\Omega &=\left( \frac{\partial \mathcal{M}}{\partial \omega }\right)
_{S,Q,P,g_{s}}=-\frac{3\sigma \ln \left( \sqrt{\frac{2S}{\pi }}\right) }{%
	4\left( \frac{2S}{\pi }\right) ^{3\omega /2}},  \label{OmSQPb}
\end{align}
}
where $\mathfrak{F}_{2_{S}}=\left. \mathfrak{F}_{2_{+}}\right\vert _{r_{+}=%
	\sqrt{\frac{2S}{\pi }}}=_{2}F_{1}\left( \left[ \frac{1}{2},\frac{3}{4}\right]
,\left[ \frac{7}{4}\right] ,-\frac{4S^{2}}{\pi ^{2}l_{s}^{4}}\right) $.

Within the extended phase space, the first law of thermodynamics is
satisfied by the conserved and thermodynamic quantities presented in Eqs. (%
\ref{MSQPb})-(\ref{OmSQPb}). This relationship is characterized as follows 
\begin{equation}
d\mathcal{M}=TdS+\Phi dQ+VdP+\mathcal{G}dg_{s}+\mathcal{L}dl_{s}+\Sigma
d\sigma +\Omega d\omega .
\end{equation}

By examining the scaling behavior of the thermodynamic variables under a uniform spatial rescaling, we establish that the mass function behaves as a weighted-homogeneous function of degree one. Applying Euler's theorem to this homogeneous relation yields a differential scaling identity. Substituting the conjugate thermodynamic variables into this identity leads to the modified Smarr formula, which takes the form
\begin{equation}
\mathcal{M}=2\left( TS-PV\right) +\Phi Q+\mathcal{L}l_{s}+\left( 3\omega +1\right) \Sigma \sigma.
\end{equation}
This formulation explicitly incorporates the extra work contributions originating from both the quintessence field and the Letelier--Alencar cloud of strings.

\section{Photon Cylinder}

\label{Photon Cylinder}

In this section, we investigate the null geodesic structure of the charged
Letelier--Alencar black string. In spherically symmetric spacetimes, the
existence of circular photon orbits gives rise to the well-known photon
sphere, which plays a crucial role in gravitational lensing and the shadow
of black holes. For cylindrically symmetric spacetimes, the analogous
structure is a photon cylinder, consisting of null circular orbits at a
fixed radial coordinate. We begin by considering the Lagrangian for a test
particle moving in the spacetime described by the metric (\ref{metric}): 
\begin{equation}
\mathcal{L} = \frac{1}{2}g_{\mu\nu}\dot{x}^{\mu}\dot{x}^{\nu},
\label{Lag_null}
\end{equation}
where the dot denotes differentiation with respect to an affine parameter.
For null geodesics, which describe the trajectories of massless particles
such as photons, we have $\mathcal{L} = 0$. Substituting the metric
coefficients from Eq.~(\ref{metric}), this condition yields 
\begin{equation}
-f(r)\dot{t}^{2} + \frac{\dot{r}^{2}}{f(r)} + r^{2}\dot{\phi}^{2} +
\alpha^{2}r^{2}\dot{z}^{2} = 0.  \label{null_metric}
\end{equation}

The spacetime under consideration possesses three Killing vectors, namely $%
\partial_t$, $\partial_\phi$, and $\partial_z$, corresponding to the
invariance under time translations, rotations around the symmetry axis, and
translations along the $z$-direction, respectively. These symmetries give
rise to three conserved quantities along the geodesic: 
\begin{equation}
\dot{t} = \frac{E}{f(r)}, \qquad \dot{\phi} = \frac{L}{r^{2}}, \qquad \dot{z}
= \frac{P}{\alpha^{2}r^{2}},  \label{conserved}
\end{equation}
where $E$ is the conserved energy, $L$ is the conserved angular momentum
about the symmetry axis, and $P$ is the conserved linear momentum along the $%
z$-direction. These constants of motion are essential for reducing the
geodesic equations to an effective one-dimensional problem.

Substituting these expressions into Eq.~(\ref{null_metric}) and rearranging
the terms, we obtain the radial equation of motion for null geodesics: 
\begin{equation}
\dot{r}^{2} = E^{2} - V_{\text{eff}}(r),  \label{rdot_lambda}
\end{equation}
where we define the quantity $\lambda^{2} \equiv L^{2} + P^{2} /\alpha^2$,
which combines the two conserved quantities associated with the Killing
vectors $\partial_\phi$ and $\partial_z$ into a single effective constant.
With this definition, the effective potential is given by 
\begin{equation}
V_{\text{eff}}(r) \equiv \frac{\lambda^{2} f(r)}{r^{2}}.  \label{eff_pot}
\end{equation}

Equation~(\ref{rdot_lambda}) has a clear mechanical interpretation: the
radial motion of a photon is governed by the effective potential $V_{\text{%
eff}}(r)$, with $E^{2}$ playing the role of the total energy. Photons can
only access regions where $E^{2} \geq V_{\text{eff}}(r)$, and turning points
occur when $E^{2} = V_{\text{eff}}(r)$.

The shape of $V_{\text{eff}}(r)$ depends crucially on the metric function $%
f(r)$, which in turn contains all the parameters of the model: the mass
parameter $m_0$, the electric charge $q$, the cosmological constant $\Lambda$%
, the quintessence parameters $\sigma$ and $\omega$, and the
Letelier--Alencar cloud of strings parameters $g_s$ and $l_s$.

Circular null orbits, which form the photon cylinder, exist when a photon
can remain at a fixed radial coordinate $r$ throughout its motion. This
requires the following two conditions to be satisfied simultaneously: 
\begin{equation}
V_{\text{eff}}(r) = E^{2}, \qquad V_{\text{eff}}^{\prime }(r) = 0.
\label{circular_conditions}
\end{equation}

The first condition ensures that the radial velocity $\dot{r}$ vanishes at
the orbit, while the second condition guarantees that the orbit is an
extremum of the effective potential. The condition $V_{\text{eff}}^{\prime
}(r) = 0$ yields 
\begin{equation}
\left.\frac{d}{dr}\left(\frac{f\left(r\right)}{r^{2}}\right)\right|_{r=r_{%
\text{pc}}}=0.  \label{pot_deriv}
\end{equation}

Using the metric function from Eq. (\ref{f(r)}) and its derivative, and
substituting into Eq.~(\ref{pot_deriv}), we obtain the transcendental
equation that determines the photon cylinder radius: 
\begin{equation}
\begin{aligned}
&\frac{12m_{0}}{r_{\text{pc}}}-\frac{16q^{2}}{r_{\text{pc}}^{2}}-\frac{%
3\left(\omega+1\right)\sigma}{r_{\text{pc}}^{3\omega+1}}\\
&\quad-\frac{g_s^2l_s^2\sqrt{1+\frac{r_{\text{pc}}^4}{l_s^4}}}{r_{\text{pc}}^2}-\frac{3g_s^2l_s^2\mathfrak{F}_{1_{\text{pc}}}}{r_{\text{pc}}^2}=0
\end{aligned}
\label{photon_eq_final}
\end{equation}

\begin{table*}[t]
\caption{Values of the first, $r_{\mathrm{pc},1}$, and second, $r_{\mathrm{pc},2}$, photon-cylinder radii for different values of the string length $l_s$, with the effective coupling fixed at $g_s=0.5$ (left panel), and for different values of $g_s$, with the string length fixed at $l_s=0.5$ (right panel). The remaining parameters are fixed at $m_0=0.9$, $q=0.3$, $\sigma=0.1$, and $\omega=-2/3$.}
\begin{center}
\begin{tabular}{  c  c  c || c  c  c  }
\hline
\hline
 $l_s$ & $r_{\text{pc},1}$ & $r_{\text{pc},2}$ & $g_s$ & $r_{\text{pc},1}$ & $r_{\text{pc},2}$\\
 \hline
 \quad 0.0 \quad & \quad 0.133 \quad & \quad 13.1 \quad &  \quad 0.0 \quad & \quad 0.133 \quad & \quad 10.3 \quad  \\
 \quad 0.5 \quad & \quad 0.157 \quad & \quad 12.8 \quad &  \quad 0.5 \quad & \quad 0.157 \quad & \quad 12.8 \quad  \\
 \quad 1.0 \quad & \quad 0.226 \quad & \quad 12.5 \quad &  \quad 1.0 \quad & \quad 0.226 \quad & \quad 23.5 \quad  \\
 \quad 1.5 \quad & \quad 0.342 \quad & \quad 12.2 \quad &  \quad 1.5 \quad & \quad 0.342 \quad & \quad 46.1 \quad  \\
 \quad 2.0 \quad & \quad 0.505 \quad & \quad 11.9 \quad &  \quad 2.0 \quad & \quad 0.501 \quad & \quad 80.1 \quad  \\
 \hline
 \hline
\end{tabular}
\label{T}
\end{center}
\end{table*}

In the limit $g_s \to 0$ and $\sigma \to 0$, this reduces to the known
result for the charged static black string. For a neutral configuration ($q
= 0$) without the extra matter fields, the equation has no finite solution,
indicating the absence of a photon cylinder, as expected. The presence of
the Letelier--Alencar cloud of strings and quintessence introduces new terms
that can generate photon cylinder solutions even for neutral configurations. In Table \ref{T} we show a detailed numerical analysis of Eq. (\ref{photon_eq_final}), determining the existence and location of photon cylinders as functions of the model parameters. On the left we show the values of the first, $r_{\mathrm{pc},1}$, and second, $r_{\mathrm{pc},2}$, photon-cylinder radii as a function of the string length $l_s$. We can conclude that first photon radius increases with the increasing the length of the string. For the second radius $r_{\mathrm{pc},2}$, the opposite occurs, that is, $r_{\mathrm{pc},2}$ decreases when $l_s$ increases. On the right side of the table we show how the two radii change when we fix $l_s$ and vary the value of the effective coupling $g_s$, we can conclude that both the first and second radius increase with the increasing of the coupling.

\section{Conclusions}
\label{Conclusions}
In this work, we have derived charged black string solutions
in Einstein--Maxwell--$\Lambda$ gravity coupled to a Letelier--Alencar cloud of strings and a Kiselev anisotropic fluid describing quintessence. The obtained solution contains several previously known black string geometries as particular cases, including the charged black string of Lemos and Zanchin, as well as the limits corresponding to the original Letelier cloud of strings and the absence of quintessence.

The geometrical structure of the spacetime was characterized by evaluating the Ricci and Kretschmann scalars. The Kretschmann scalar and the Ricci scalar reveal the existence of an essential curvature singularity located at the symmetry axis. We subsequently explored the thermodynamic aspects of the obtained solution by deriving its conserved and thermodynamic quantities, including the Hawking temperature, entropy, electric charge, electric potential, and mass. 

The local and global thermodynamic stability properties were then examined through the heat capacity and Gibbs potential, respectively. In particular, we analyzed the influence of the relevant physical parameters on the corresponding stability domains. Finally, by promoting the cosmological constant to a thermodynamic pressure, we extended the thermodynamic framework and derived the associated extended first law together with the Smarr relation.

The null geodesic structure was investigated by deriving the equation that
determines the photon cylinder radius. The resulting expression extends the
corresponding equation for charged black strings by incorporating the
contributions of the Letelier--Alencar cloud of strings and the Kiselev
fluid. For $\sigma>0$ and a simple outer horizon, the solution
admits an exterior photon cylinder. The charged black string of Barbosa
and da Silva~\cite{Barbosa2026arXiv260606435B}, with an ordinary Letelier cloud
and a Kiselev fluid, has no exterior photon cylinder in the regime they studied.

A natural continuation of this work is the extension of these solutions to
the rotating case. Rotating black strings exhibit a richer causal and
thermodynamic structure, and it would be interesting to examine how the
generalized cloud of strings and the quintessence field modify their
geometry, conserved quantities, thermodynamic behavior, and null geodesic
structure. Such an extension may also provide further insight into the role
of anisotropic matter distributions in cylindrically symmetric rotating
spacetimes.

\begin{acknowledgments}
This research was financed by a research grant from the University of Mazandaran. The
authors also acknowledge the use of AI tools only for language polishing and
improving the clarity of the manuscript. L.G.B. acknowledges the financial support of the Coordenação de Aperfeiçoamento de Pessoal de Nível Superior (CAPES), Brazil (Finance Code 001).
\end{acknowledgments}















\appendix
\section{Comparison of static charged black string solutions}
\label{app:comparison}

Table~\ref{tab:comparison} compares the two cases covered by
Eq.~(\ref{f(r)}) with the charged black string of Lemos and Zanchin and
the charged black string with a Kiselev fluid and an ordinary Letelier
cloud. The same mass and charge normalization is used throughout.
For fixed $r>0$, the Letelier--Alencar term tends to $-g_s^2$ as
$l_s\to0$, up to a mass term absorbed into $m_0$.

\begin{table}[!ht]
\caption{Matter sources for static charged black strings in Einstein--Maxwell--$\Lambda$ theory.}
\label{tab:comparison}
\centering
\begin{ruledtabular}
\begin{tabular}{p{.38\columnwidth}p{.51\columnwidth}}
Solution & {\raggedright Sources in addition to the Maxwell field\par} \\
\hline
Lemos and Zanchin \cite{Lemos:1995cm} & None \\[3pt]
{\raggedright Barbosa and da Silva~\cite{Barbosa2026arXiv260606435B}\par} &
Letelier cloud and Kiselev fluid \\[3pt]
This work, $\sigma=0$ &
Letelier--Alencar cloud \\[3pt]
This work, $\sigma\neq0$ &
{\raggedright Letelier--Alencar cloud and Kiselev fluid\par} \\[3pt]
\end{tabular}
\end{ruledtabular}

\end{table}

\bibliography{reference}

\end{document}